\documentstyle[aps,epsfig]{revtex}
\begin{document}
\title{Stability of self-consistent solutions for the Hubbard model at
  intermediate and strong coupling}

\author{V.  Jani\v{s}}

\address{Institute of Physics, Academy of Sciences of the Czech Republic,\\
  Na Slovance 2, CZ-18221 Praha 8, Czech Republic\\ e-mail:janis@fzu.cz}

\date{\today}
\maketitle 
\begin{abstract}
  We present a general framework how to investigate stability of solutions
  within a single self-consistent renormalization scheme being a
  parquet-type extension of the Baym-Kadanoff construction of conserving
  approximations. To obtain a consistent description of one- and
  two-particle quantities, needed for the stability analysis, we impose
  equations of motion on the one- as well on the two-particle Green
  functions simultaneously and introduce approximations in their input, the
  completely irreducible two-particle vertex. Thereby we do not loose
  singularities caused by multiple two-particle scatterings. We find a
  complete set of stability criteria and show that each instability,
  singularity in a two-particle function, is connected with a
  symmetry-breaking order parameter, either of density type or anomalous.
  We explicitly study the Hubbard model at intermediate coupling and
  demonstrate that approximations with static vertices get unstable before
  a long-range order or a metal-insulator transition can be reached. We use
  the parquet approximation and turn it to a workable scheme with dynamical
  vertex corrections. We derive a qualitatively new theory with
  two-particle self-consistence, the complexity of which is comparable with
  FLEX-type approximations. We show that it is the simplest consistent and
  stable theory being able to describe qualitatively correctly quantum
  critical points and the transition from weak to strong coupling in
  correlated electron systems.
\end{abstract}


\section{Introduction}
\label{sec:intro}

One of the major problems in the theory correlated electrons is to
construct in a systematic and controlled way a consistent approximation
interpolating reliably between weak- and strong-coupling regimes.  The two
extreme limits of weak and strong couplings in the archetypal Hubbard model
can be described relatively well. The weak-coupling regime is governed by a
Hartree-Fock mean field with dynamical fluctuations covered by Fermi-liquid
theory. Extended systems at low temperatures are Pauli paramagnets with
smeared out local magnetic moments. For bipartite lattices
antiferromagnetic long-range order sets in at half filling and zero
temperature at arbitrarily small interaction.  In the strong-coupling
regime the Hubbard model at half filling maps onto a Heisenberg
antiferromagnet with pronounced local magnetic moments and the Curie-Weiss
law for the staggered magnetic susceptibility, at least at the mean-field
level. The spectral structure is dominated by separated lower and upper
Hubbard bands and the strongly correlated system seems to be insulating
even in the paramagnetic phase.

However, it is the intermediate coupling, where the effective Coulomb
repulsion is comparable with the kinetic energy and hence neither very weak
nor very strong, that is of great interest for the theorists as well as for
the experimentalists.  At intermediate coupling, dynamical fluctuations
control the low-temperature physics of interacting electrons and neither
weak-coupling nor atomic-like perturbation theories are adequate.  In this
nonperturbative regime a singularity in a two-particle function is
approached and we expect breakdown of the Fermi-liquid regime and a
transition to an ordered state breaking the symmetry of the Hamiltonian.
Most of the theoretical effort in strongly correlated electron systems
concentrates on finding an appropriate and reliable description of this
fluctuation-dominated transition region.

There are only a few general theoretical means with which we could address
the weak-to-strong-coupling transition. The only exact, Bethe-ansatz
solution in $d=1$ \cite{Lieb68} does not have a weak-coupling regime and
numerical techniques such as quantum Monte-Carlo simulations or exact
diagonalization are restricted to relatively high temperatures or small
clusters \cite{Dagotto94}. They cannot cope efficiently with the very
different energy scales relevant in the transition regime. Recent progress
in the dynamical mean-field theory constructed from the limit $d\to\infty$,
where only frequency fluctuations are dynamically relevant, has brought new
insights into possible scenarios of the transition from weak to strong
couplings.  The transition at the mean-field level is connected with a
Kondo-like behavior and the Mott-Hubbard metal-insulator transition. Its
critical behavior has not yet been fully cleared.  A widely accepted
picture for the transition at zero temperature, initially based on the
so-called iterated perturbation theory (IPT) \cite{Zhang93} and lately
supported by analytical \cite{Moeller95} and numerical \cite{Bulla98}
renormalization-group arguments, was, however, recently questioned using
assessments from perturbation theory and a skeleton expansion
\cite{Kehrein98}.  Unfortunately even in this particular limit of high
spatial dimensions, no exact solution to the Hubbard model has yet been
found and our predictions and conclusions are justified by approximate
schemes of either numerical or analytical character \cite{Georges96}.  The
need for a nonperturbative quantitative scheme being stable throughout the
whole range of the interaction strength is still extreme.

We concentrate in this paper exclusively on analytically controllable
approximations based on partial summations and renormalizations of Feynman
diagrams. Only these approaches have the power to give the details of the
critical behavior of correlated electrons.  To decide which of the existing
approaches are prospective in a reliable description of quantum critical
behavior, we investigate in detail stability of approximations constructed
within renormalized weak-coupling expansions continued to intermediate and
strong couplings.

We use the framework of Baym and Kadanoff \cite{Baym61,Baym62} to formulate
general stability criteria that must be imposed upon any approximate theory
intending to be thermodynamically consistent and conserving. I.~e. we use
an expansion with renormalized one-electron propagators and construct a
Luttinger-Ward generating functional from which all the physical quantities
are derived via functional derivatives.  Additively we use small external
perturbations to test  stability of the equilibrium states.

In this paper we extend the Baym and Kadanoff scheme in two important
aspects. First, we generalize external perturbations of the equilibrium
system from density fluctuations also to anomalous sources not conserving
charge and/or spin. Only in this way we control all relevant two-particle
vertex functions that may show divergences and hence lead to instabilities.
We use the disturbances as generators of two-particle irreducibility and
find a complete set of criteria for a local stability of self-consistent
solutions with a rigorous relation between singularities in two-particle
functions, indicating an instability of the solution, and symmetry-breaking
order parameters.
 
Second, since two-particle functions play extremely important
role in the stability analysis we have to treat the vertex functions and
the one-electron propagators on the same footing within a single
renormalization scheme. We must not loose any important relation between
various two-particle quantities as well as between one- and two-particle
functions. This is achieved if instead of approximating the self-energy
functional we shift  diagrammatic approximations to the level of
two-particle functions.

Equations of motion for one- and two-particle Green functions are essential
in our construction of approximate theories. They are the Schwinger-Dyson
and Bethe-Salpeter equations.  The former connects the one-particle
self-energy with the full two-particle vertex function. The latter
equations are defined separately for each two-particle irreducibility
channel and determine finally the full vertex as a functional of the
completely two-particle irreducible one.  Since we expect that the
Bethe-Salpeter equations may get singular and the full vertex function can
have poles at intermediate and strong coupling, we do not attempt to
destroy the structure of these equations of motion.  Instead we introduce
diagrammatic approximations only in the input to these equations, i.~e. in
the completely irreducible two-particle vertex.  Thereby we do not miss
possible singularities of the vertex function assumed the relevant physics
is essentially determined by two-particle scatterings. Such a
renormalization of Feynman diagrams, or construction of skeleton diagrams,
called parquet approach \cite{Dominicis62,Jackson82}, is from the very
beginning self-consistent at the level of one- as well as two-particle
functions.  It contains dynamical vertex renormalizations and guarantees
that divergences in two-particle functions are treated properly.

The parquet approach or the parquet diagrams are known for long in
many-body physics. First introduced in nuclear physics \cite{Sudakov56},
the parquet theory found quickly its way into condensed matter. It was
attempted to solve the Kondo problem \cite{Abrikosov64}, the x-ray edge
\cite{Roulet69}, or the formation of the local magnetic moment in dilute
alloys \cite{Weiner70} with the parquet approximation. However, in none of
these applications a full solution to the parquet equations was found. In
the former two approaches an expansion with leading-logarithmic divergences
of single electron-hole bubbles was used.  This construction underestimates
the role of the genuine poles in the Bethe-Salpeter equations due to
multiple two-particle scatterings.  It actually amounts to a loop expansion
neglecting the complex dynamics of the vertex functions.  The last approach
resorted to static approximations and had only a restricted success in the
strong-coupling limit.  It is the intricate structure of the parquet
equations that makes nonperturbative solutions inavailable and hinders a
broader application of the parquet construction.

Having a general scheme for deriving theories with renormalized one- and
two-particle functions we must reach for approximations to come to
quantitative results. The simplest theory fulfilling the equations of
motion for both one- and two-particle Green functions is the parquet
approximation where the completely irreducible vertex is just the bare
interaction.  However, up to now no nonperturbative solution to the parquet
equations has yet been found. There already exist simplifications of the
parquet construction (vertex renormalization) \cite{Bickers91,Vilk94}, but
because they resort to static vertices they get unstable and fail at strong
coupling.  Here we reduce the intricate complex structure of the full
parquet approximation so that the vertex functions are dynamically
renormalized and allow only for integrable divergences. We demand an
asymptotic accuracy of the simplification at critical points and keep only
maximally divergent contributions to the vertex function
\cite{Janis98a,Janis99a}. We end up at zero temperature and half filling
with a theory stable throughout the whole range of the interaction strength
the computational complexity of which is comparable with the single-channel
or FLEX-type approximations. We apply it to a mean-field description of the 
Mott-Hubbard metal insulator transition.

The layout of the paper is as follows. Section~\ref{sec:stability} presents
the general scheme for deriving approximations with renormalized one- and
two-particle propagators allowing us to formulate stability criteria with
the relevant two-particle functions. We explicitly show in
Sections~\ref{sec:hartree},~\ref{sec:second-order} that all approximations
simpler than the parquet diagrams get unstable at intermediate coupling of
the Hubbard and single-impurity Anderson models. Stability criteria for the
parquet approximation are explicitly formulated in Sec.~\ref{sec:parquet}.
The proposed simplification of the parquet approximation with dynamical
vertex corrections is presented in Sec.~\ref{sec:simplifications}. We
demonstrate  stability of the simplified parquet approach in the
mean-field theory of the Mott-Hubbard metal-insulator transition in
Sec.~\ref{sec:DMFT}. Last Section~\ref{sec:conclusions} brings discussion
and concluding remarks.

\section{Equations of motion, two-particle Green functions and
  stability of solutions}
\label{sec:stability}

We use two generic Hubbard and single-impurity Anderson Hamiltonians with
an external magnetic field $B$ to model moderate and strong electron
correlations. The magnetic field allows us to lift the degeneracy in the
spin space and to determine the relevant two-particle functions we need to
look at if we want to prove  stability of the theory. We start
with equilibrium Hamiltonians
\begin{mathletters} 
\begin{eqnarray}
  \label{eq:hh}
  \widehat{H}_H&=&\sum_{{\bf k}\sigma} \left(\epsilon({\bf k})
    +\mu+\sigma B\right) c^{\dagger}_{{\bf k}\sigma}
  c^{\phantom{\dagger}}_{{\bf k}\sigma}  +
  U\sum_{{\bf i}}\widehat{n}_{{\bf i}\uparrow}\widehat{n}_{{\bf i}
    \downarrow},  \\
 \label{eq:ah}
 \widehat{H}_A&=&\sum_{{\bf k}\sigma} \left(\epsilon({\bf k})
   +\sigma B\right)
 c^{\dagger}_{{\bf k}\sigma}c^{\phantom{\dagger}}_{{\bf k}\sigma}
 +\sum_{{\bf k}\sigma}  
 \left(V_{{\bf k}} c^{\dagger}_{{\bf k}\sigma}f_\sigma +V_{{\bf k}}^*
   f^\dagger_\sigma c^{\phantom{\dagger}}_{{\bf k}\sigma}\right)
 +\sum_\sigma(\varepsilon_f +\sigma B) f^\dagger_\sigma f_\sigma
 +U\widehat{n}_{f\uparrow}\widehat{n}_{f\downarrow} .  
\end{eqnarray}
\end{mathletters}
The Hubbard Hamiltonian (\ref{eq:hh}) is of primary interest for us, since
phase transitions with long-range order can appear there. The Anderson
Hamiltonian (\ref{eq:ah}) is considered only to demonstrate the difference
between a mean-field solution of the Hubbard model ($d=\infty$) and the
impurity solution and to test adequacy of chosen approximate schemes.

The most direct way to study stability of solutions at thermal equilibrium
is to introduce auxiliary sources as an external perturbation
\cite{Baym61}.  Only external sources conserving spin and charge leading to
density-density (nonanomalous) response functions are standardly used to
perturb the equilibrium state. However, these sources are not complete and
do not allow for anomalous responses and quantum phases such as
superconductivity. To be able to describe all possible instabilities of the
Hubbard-like models we introduce a generalized external perturbation
\begin{eqnarray}
  \label{eq:H-ext}
  \widehat{H}_{ext}&=&\int
      d1d2\Bigg\{\sum_\sigma\left[\eta^{||}_\sigma(1,2) c^{\dagger}
      _{\sigma}(1) c^{\phantom{\dagger}}_{\sigma}(2)+\bar{\xi}^{||}_\sigma
      (1,2)c^{\phantom{\dagger}}_{\sigma}(1)c^{\phantom{\dagger}}_{\sigma}
      (2)+\xi^{||}_\sigma(1,2) c^{\dagger}_{\sigma}(1)c^{\dagger}
      _{\sigma}(2)\right]\nonumber\\ 
  &&+\left[\eta^{\perp}(1,2)c^{\dagger}_{\uparrow}(1)c^{\phantom{\dagger}}
      _{\downarrow}(2) +\bar{\eta}^{\perp}(1,2)c^{\dagger}_{\downarrow}(2)
      c^{\phantom{\dagger}}_{\uparrow}(1)\right]+\left[\bar{\xi}^{\perp}
      (1,2)c^{\phantom{\dagger}}_{\uparrow}(1)c^{\phantom{\dagger}}
      _\downarrow(2)+\xi^{\perp}(1,2)c^{\dagger}_{\downarrow}(2)
      c^{\dagger}_{\uparrow}(1)\right]\Bigg\}
\end{eqnarray}
where labels $1=({\bf r}_1,t_1)$, $2=({\bf r}_2,t_2)$ denote a set of
space-time variables characterizing the motion of quantum particles,
electrons.  Here the local real fields $\eta^{||}_\sigma$ induce density
responses conserving charge as well as spin. These fields (e.~g. magnetic)
are used in standard stability analyses.  The new nonconserving external
sources are complex and either add or remove charge, spin or both from the
equilibrium many-particle state.  The field $\eta^{\perp}$ conserves charge
but increases spin of the equilibrium state by two elementary units. The
fields $\xi^{\perp}$ increase charge while $\xi^{||}_\sigma$ increase both
charge and the appropriate spin projection.  The complex conjugate fields
lower the respective quantities.

Fundamental quantity for quantum statistics is a thermodynamic potential as
a functional of the external perturbation $\widehat{H}_{ext}$. We need to
expand the generating thermodynamic potential only to second order in the
external perturbation in order to determine (local) stability of the
equilibrium state.  First term of this expansion generates one-particle
Green functions. The real fields lead to regular propagators, while the
complex ones to anomalous Green functions. They vanish at equilibrium
unless we are in a quantum phase with anomalous Green functions. To decide
whether a long-range order may arise or whether the equilibrium state is
stable we have to evaluate second term in the external perturbation. It
defines two-particle functions, regular and anomalous. In the unperturbed,
equilibrium state without anomalous functions only diagonal terms in the
external sources do not vanish. If $H_\alpha$ denotes a chosen external
source perturbing the equilibrium system, i.~e. it stands generically for
$\eta^{||}$, $\eta^{\perp}$, $\xi^{||}$, and $\xi^{\perp}$, we can write
\begin{eqnarray}
  \label{eq:G2}
  G^{(2)\alpha}(13,24)&=&\frac{\delta^2\Phi[G,H]}{\delta H_{\alpha}(4,3)
  \delta H_{\bar{\alpha}}(2,1)}\Bigg|_{H=0}\ . 
\end{eqnarray}
It is interesting to note that second derivatives w.r.t. different external
sources introduced in perturbation $\widehat{H}_{ext}$, (\ref{eq:H-ext}),
lead at equilibrium to inequivalent two-particle irreducibility channels,
diagrams that cannot be made disconnected by cutting two specific lines.
More precisely, the external sources generate two-particle reducible
functions, while their Legendre conjugate ``order parameters'' the
respective irreducible functions.  The field $\eta^{||}$ leads to reducible
functions in the interaction ($U$) channel and evokes normal density
responses and susceptibilities. The field $\eta^{\perp}$ leads
to the electron-hole ($eh$) and $\xi^{\perp}$ to the electron-electron
($ee$) reducible functions, respectively.  Second derivative w.r.t.
$\xi^{||}$ vanishes at equilibrium without anomalous functions \cite{note0}.

It is more convenient to introduce a vertex function instead of the full
two-particle function $G^{(2)\alpha}$. We subtract the free two-particle
propagation, if contributes, and stripe off the external legs from the
remaining two-particle function. The result is a vertex function $\Gamma$
independent of the index $\alpha$ \cite{note1}. The full vertex function
$\Gamma$ can be decomposed equivalently in each two-particle channel into a
sum of the irreducible and reducible functions $\Lambda$ and $\mathcal{K}$,
respectively:
\begin{eqnarray}
  \label{eq:vertex-def}
  \Gamma&=&\Lambda^\alpha+\mathcal{K}^\alpha .
\end{eqnarray}

\begin{figure}
  \hspace*{80pt} \epsfig{figure=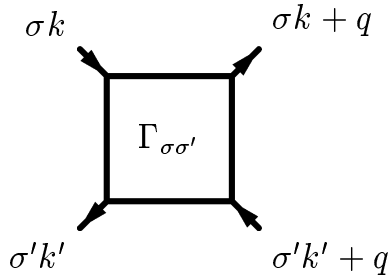,height=40mm}
\caption{\label{fig:2P-generic} Generic two-particle function with three
  independent four-momenta and a defined order of incoming and outgoing
  fermions.}
\end{figure}

To go on we have to introduce a convention for independent variables
labeling the two-particle functions. We denote $X_{\sigma\sigma'}(k,k';q)$
the generic (regular) two-particle function in the momentum representation.
We use four-momenta, $k=({\bf k},i\omega_n)$ for the fermionic and $q=({\bf
  q},i\nu_m)$ for the bosonic variables with Matsubara frequencies
$\omega_n=(2n+1)\pi$, $\nu_m=2m\pi$.  The meaning of each variable in the
function $X$ is manifested in Fig.~\ref{fig:2P-generic}.  The incoming
arrows denote annihilation of charge.

A two-particle irreducible function in a specified channel $\alpha$ can be
defined as a functional derivative of the self-energy $\Sigma^{\alpha}$
from the $\alpha$-channel with respect to the renormalized one-electron
propagator
\begin{mathletters}\label{eq:IGF-diff}
\begin{eqnarray}
  \label{eq:2IP-diff}
  \Lambda^\alpha(13,24)&=&\frac{\delta\Sigma^\alpha(1,2)}{\delta
    G^\alpha(4,3)} \Bigg|_{H=0} \ ,
\end{eqnarray}
where the self-energy is a functional of the full one-electron propagator
defined from the generating Luttinger-Ward functional
\begin{eqnarray}
  \label{eq:1IP-diff}
  \Sigma^\alpha(1,2)&=&\frac{\delta\Phi[G,H]}{\delta
    G^\alpha(2,1)}\ . 
\end{eqnarray}\end{mathletters}
Note that even if the anomalous self-energy $\Sigma^{\alpha}$ vanishes at
equilibrium the two-particle function is nontrivial, since it is conserves
both charge and spin.

Up to now we looked at the equilibrium system from the thermodynamic point
of view. We assumed we have a Luttinger-Ward generating functional
$\Phi[G]$ from which we determine all the relevant quantities via
functional derivatives. However, we do not know how to find the
Luttinger-Ward functional. A naive way is to choose a particular set of
skeleton diagrams directly for the grand potential. A more convenient and
usual way is to choose diagrams at the one-particle level by fixing the
functional $\Sigma[G]$. However, in neither of these constructions we have
a connection to the fundamental equations of motion for the one- and
two-particle Green functions. The equations of motion determine the
microscopic dynamics of quantum systems. If we want to simulate the exact
dynamics of the studied system we must try to approximate the equations of
motion as accurate as possible.

First equation of motion we have to fulfill is the Schwinger-Dyson equation
connecting the one-electron self-energy with the vertex function. It reads
for Hubbard-like models with a local electron-electron interaction:
\begin{eqnarray}
  \label{eq:sigma-2P}
  \Sigma_\sigma(k)&=&\frac{U}{\beta N}\sum_{k'} G_{-\sigma}(k') -\frac{U}
  {\beta^2N^2}\sum_{k'q}\Gamma_{\sigma-\sigma}(k,k';q)
  G_\sigma(k+q)G_{-\sigma}(k'+q)G_{-\sigma}(k').     
\end{eqnarray}
The first term on the right-hand side is the Hartree self-energy and the
latter contains second and higher-order, dynamical corrections.  The usual
way to deal with the Schwinger-Dyson equation is to introduce
approximations to the vertex function $\Gamma[G]$ on the right-hand side of
(\ref{eq:sigma-2P}). We thereby get a closed functional $\Sigma[G]$ and
close the approximation \cite{Baym62}. However, at the critical points,
where equilibrium states get unstable, the full vertex function $\Gamma$ is
singular. Brute interferences in the Schwinger-Dyson equation can
make the approximate dynamics inadequate and qualitatively different from
the exact one. Since we are primarily interested in the two-particle
functions and their appropriate description, we move approximations from
the full vertex function to a lower level, namely to the completely
irreducible two-particle function.

Two-particle irreducible vertices determine the full vertex function via
Bethe-Salpeter equations.  These equations explicitly sum two-particle
reducible diagrams in each two-particle channel.  In the notation of
Fig.~\ref{fig:2P-generic} we obtain
\begin{mathletters}\label{eq:Bethe-Salpeter}
\begin{eqnarray}
  \label{eq:2P-reducible}
  \Gamma(k,k';q)&=&\Lambda^\alpha(k,k';q)-\left[\Lambda^\alpha GG\odot
    \Gamma \right](k,k';q) . 
\end{eqnarray}  
The Bethe-Salpeter equations must be completed by definitions connecting
the respective irreducible and reducible functions $\Lambda^\alpha$, ${\cal
  K}^\alpha$, respectively.  Due to inequivalence of different two-particle
channels we have
\begin{eqnarray}
  \label{eq:2IP-def}
  \Lambda^\alpha&=& I+\sum_{\alpha'\neq\alpha}{\cal K}^{\alpha'}
\end{eqnarray}
\end{mathletters}
where $I[U;G,\Gamma]$ is the completely irreducible two-particle vertex
being a functional of the one-particle propagator $G$ and generally also of
the vertex function $\Gamma$.  When expressed diagrammatically, the
completely irreducible vertex can be disconnected only by cutting at least
three one-electron propagators.

We used a generic notation $\odot$ for the channel-dependent multiplication
of two-particle functions. It mixes the variables of two-particle functions
in different manners.  The three matrix multiplication schemes for
two-particle quantities in our notation of Fig.~\ref{fig:2P-generic},
representing summations over intermediate states, read in the interaction,
electron-hole, and electron-electron channels, respectively
\begin{mathletters}\label{eq:conv} 
\begin{eqnarray}
  \label{eq:conv-U}
  \left[\widehat{X}GG\star\widehat{Y}\right]_{\sigma\sigma'}(k,k';q)&=&\frac
  1{\beta{\cal N}}\sum_{\sigma''k''}X_{\sigma\sigma''}(k,k'';q)
  G_{\sigma''} (k'')G_{\sigma''}(k''+q)Y_{\sigma'' \sigma'}(k'',k';q), \\
  \label{eq:conv-eh}
  \left[\widehat{X}GG\bullet\widehat{Y}\right]_{\sigma\sigma'}(k,k';q)&=&
  \frac 1{\beta{\cal N}}\sum_{q''} X_{\sigma\sigma'}(k,k';q'')G_{\sigma}
  (k+q'')G_{\sigma'}(k'+q'')\nonumber \\&&
  \hspace*{40pt}\times Y_{\sigma\sigma'}(k+q'',k'+q'';q-q''),\\  
  \label{eq:conv-ee}
  \left[\widehat{X}GG\circ\widehat{Y}\right]_{\sigma\sigma'}(k,k';q)&=&\frac
  1{\beta{\cal N}}\sum_{q''} X_{\sigma\sigma'}(k,k'+q'';q-q'')G_{\sigma}
  (k+q-q'') G_{\sigma'}(k'+q'')\nonumber\\ &&
  \hspace*{40pt}\times Y_{\sigma\sigma'}(k+q-q'',k';q'') .
\end{eqnarray} \end{mathletters}
For convenience we used separate symbols for each multiplication scheme.
Note that only the interaction (vertical) channel mixes the spin singlet
and triplet functions.

Now all the quantities depend explicitly only on the model input parameters
and are functionals of the completely irreducible vertex $I[U;G,\Gamma]$ to
which approximations can be chosen. Two-particle functions are determined
self-consistently from (\ref{eq:Bethe-Salpeter}) and (\ref{eq:conv}). The
one-electron self-energy is defined from (\ref{eq:sigma-2P}).  The
generating Luttinger-Ward functional is determined from the functional
differential equation (\ref{eq:1IP-diff}). Its explicit form for the
parquet approximation with $I=U$ is derived in the Appendix.

Approximate theories break some exact relations and the thermodynamic and
dynamical definitions of the two-particle functions do not coincide. If we
enter an approximation, we have ambiguous definitions for two-particle
irreducible functions. They can be defined either from (\ref{eq:IGF-diff})
or from (\ref{eq:Bethe-Salpeter}). This is an unrecoverable feature of all
approximate theories which one has to live with. The only thing one can
demand from reliable approximations is that both the definitions of
two-particle functions lead to qualitatively the same physical behavior
(phase diagram and spectral properties). The vertex functions from the
equations of motion are responsible for spectral and analytic properties of
the approximate theory and are used to determine internal stability
(consistency) of approximations. On the other hand the differential
definition of the vertex functions is used for determining thermodynamic
behavior and stability with respect to external disturbances.  They may be
either due to external sources or due to neglected dynamical vertex
renormalizations.

The advantage of the above general formulation with equations of motion is
that we treat the one- and two-particle functions consistently on the same
footing within one renormalization scheme. This is essential for
determining stability of approximations. We now rewrite the
channel-dependent matrix multiplications so as to distinguish manifestly
active and conserved variables (parameters). We introduce new notations in
which only the active four-momenta are used as variables.  We define the
following symbols
\begin{mathletters}  \label{eq:channels}
\begin{eqnarray}
\label{eq:U}
  \left[X GG\right]^U[q](\sigma k,\sigma'k')&=&X_{\sigma
    \sigma'}(k,k';q)G_{\sigma'}(k')G_{\sigma'}(k'+q),\\
  \label{eq:eh}
  \left[X GG\right]^{eh}_{\sigma\sigma'}[q] (k,k')&=&X
  _{\sigma\sigma'}(k,k+q,k'-k) G_{\sigma}(k')G_{\sigma'}(k'+q),\\
  \label{eq:ee}
  \left[X GG\right]^{ee}_{\sigma\sigma'}[q](k,k')&=&X
  _{\sigma\sigma'}(k,k';q-k-k')G_{\sigma}(k')G_{\sigma'}(q-k') .
  \end{eqnarray}
\end{mathletters}
In (\ref{eq:channels}) the bosonic variable $q$ is always inactive in the
multiplication, or conserved during the scatterings within the chosen
two-particle channel. Each channel, however, has a different conserving
variable. It makes the parquet algebra of two-particle functions
complicated.

We can, in principle, find eigenvalues and eigenvectors for the matrices,
kernels of the Bethe-Salpeter equations (\ref{eq:Bethe-Salpeter}),
$[\Lambda^\alpha GG]^\alpha$, and denote them $Q^\alpha$. These are complex
four-vectors. Only real four-vectors, i.e. for zero frequency, both the
conserved $q^\alpha$ as well as the eigenvectors $Q^\alpha$ are important
for determining the stability conditions. They may formally be written as
\begin{mathletters} \label{eq:stability}
\begin{eqnarray}
  \label{eq:stability1}
  \mbox{min}\left[\Lambda^{\alpha}GG\right]^\alpha[\textbf{q}^\alpha,0]
  (\textbf{Q}^\alpha,0)  &\ge& -1 
\end{eqnarray}
or equivalently using (\ref{eq:2IP-def})
\begin{eqnarray}
  \label{eq:stability2}
  \mbox{min}\left[\left(I+\sum_{\alpha'\neq\alpha}{\cal K}^{\alpha'}
  \right)GG\right]^\alpha[\textbf{q}^\alpha,0](\textbf{Q}^\alpha,0)&\ge&-1 . 
\end{eqnarray}\end{mathletters}
This is a complete set of conditions for a local stability of any symmetric
solution fulfilling the Bethe-Salpeter equations of motion. The
inequalities are a direct consequence of the nonexistence of singularities
at the real axis in equations (\ref{eq:Bethe-Salpeter}). Equality in
(\ref{eq:stability}) indicates that a pole in the respective two-particle
(reducible) Green function appears and consequently a singularity in a
correlation function such as susceptibility. Due to (\ref{eq:2IP-def}), the
singularity is transferred to the irreducible functions from the other
two-particle channels and unless the singularity is {\em integrable}, the
scatterings with singular irreducible kernels destroy the transition. We
hence can have only integrable singularities in the exact theory.  A
singularity in an approximate two-particle function means on one hand the
existence of long-range correlations as a precursor of a long-range
order, but on the other hand also signals critical vertex corrections
stabilizing the symmetric phase. These two critical phenomena compete and
we must compare the symmetry breaking contributions at the critical point
with the singular vertex renormalization in approximate theories
\cite{Janis98a}.

The nonexistence of poles at the real axis is a condition for the
thermodynamic stability of the symmetric (paramagnetic) phase. However, a
consistent solution for $\Sigma$ and $\Gamma$ must produce analytic
functions in complex frequencies. This is achieved if the analytic
continuations of the two-particle functions from Matsubara frequencies have
no poles in the complex plane away from the real axis. Hence the proper
analyticity of the solution is guaranteed if
\begin{eqnarray}
  \label{eq:stabibilty3}
  \left[\Lambda^{\alpha}GG\right]^\alpha[{\bf q}^\alpha,z]({\bf Q}^\alpha,
  Z)&\neq& -1 
\end{eqnarray}
where $\mbox{Im} z$,$\mbox{Im} Z \neq 0$. If these conditions are fulfilled
the self-energy determined from (\ref{eq:sigma-2P}) is analytic in the
lower and upper complex half-planes.

The stability criteria, when applied to approximate solutions, must be used
with both two-particle functions. The two-particle functions from the
equations of motion are used to determine internal consistency of the
approximation and especially (\ref{eq:stabibilty3}) is used to prove
analyticity of the approximate Green functions. The thermodynamically
derived vertex functions are used to check stability of the approximation
with respect to fluctuations caused either by long-range pair correlations
or by neglected vertex corrections. An ``external'' instability of an
approximation hence need not automatically mean a transition to an ordered
state. It may happen for oversimplified approximations, what actually is
the case in low dimensions, that the neglected dynamical vertex
renormalization smears out the transition.  Hence, before we make
conclusions on the existence of a symmetry breaking with a long-range
order, we must be sure the neglected vertex corrections are irrelevant.  If
not, the approximate LRO becomes unreliable and further vertex
renormalizations are indispensable.

\section{Hartree approximation}
\label{sec:hartree}

We investigate in this and the following sections to what degree various
approximations produce stable equilibrium states at intermediate and strong
coupling.  Only stable solutions are reliable or suitable for the
description of the transition from weak to strong couplings.

We start with the simplest, Hartree approximation.  The Hartree
approximation is obtained from our construction if we put $\Gamma=0$ in
(\ref{eq:sigma-2P}). The theory is internally consistent in a trivial way
and the only nontrivial two-particle function is that obtained from
(\ref{eq:IGF-diff}). To determine the external stability we have to
evaluate single two-particle bubbles
\begin{eqnarray}
  \label{eq:2P-bubble}
  X_{\sigma\sigma'}({\bf q},i\nu_m)&=&\frac U{\beta{\cal N}}\sum_{{\bf k}n}
  G_\sigma({\bf k},i\omega_n) G_{\sigma'}({\bf k+q},i(\omega_n+\nu_m)) 
\end{eqnarray}
with the Hartree propagators.  They explicitly read for the Hubbard and
single-impurity models
\begin{mathletters}\label{eq:GF-def}
\begin{eqnarray}
  \label{eq:GF-Hubbard}
   G_\sigma({\bf k},z)&=&\left[z+\mu+\sigma B-Un_{-\sigma}-\epsilon({\bf
       k})\right]^{-1} , \\ \label{eq:GF-SIAM}
   G_\sigma(z)&=&\left[z-\varepsilon_f+\sigma B-Un_{-\sigma}-V^2g(z+\sigma
     B)\right]^{-1} 
\end{eqnarray}
\end{mathletters}
where $g(z)$ is the local element of the one-electron propagator of the
conduction electrons. We use $g(z)=2(z-\sqrt{z^2-1})/w^2$ corresponding to
a Bethe lattice in $d=\infty$.

\begin{figure}
  \epsfig{figure=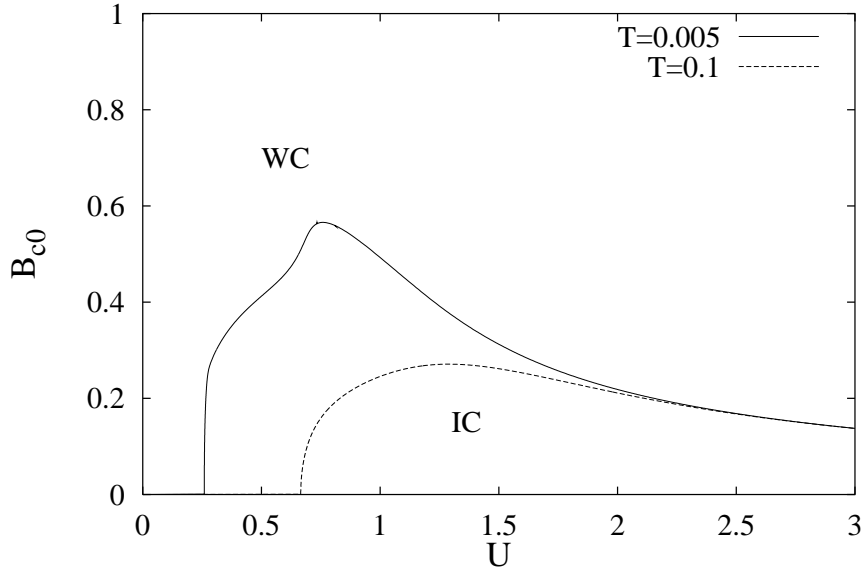,height=12cm, angle=-90}
\caption{\label{fig:hartree-d3} Instability of the Hartree approximation
  for the half-filled Hubbard model in an external magnetic field for low
  temperatures in $d=3$. The boundary field $B_{c0}$ separates weak and
  intermediate couplings (WC, IC). The energy unit is fixed by the
  half-bandwidth $w=1$.}
\end{figure}

We evaluate the bubbles for the Hubbard and single impurity models only in
the paramagnetic phase at half filling, the electron-hole symmetric
situation.  We start with arbitrary finite dimension. The singlet and
triplet electron-hole bubbles after summation over Matsubara frequencies
are:
\begin{mathletters} \label{magnetic.bubbles}
\begin{eqnarray}
  X_{\uparrow\downarrow}({\bf q},\zeta)&=&X_{\downarrow\uparrow}
  (-{\bf q},-\zeta) =\frac U{\cal N}\sum_{\bf k}\frac{f\left(\epsilon({\bf k})
    -B-Um/2\right)-f\left(\epsilon({\bf k}+{\bf q}) +B+ Um/2\right)}{\zeta
  -2B-Um +\epsilon({\bf k})-\epsilon({\bf k}+{\bf q})}, \\ 
   X_{\sigma\sigma}({\bf q},\zeta)&=&
    \frac U{\cal N}\sum_{\bf k}\frac{f\left(\epsilon({\bf k})
    -\sigma(B+ Um/2)\right)-f\left(\epsilon({\bf k}+{\bf q})-\sigma(B+Um/2)
  \right)} {\zeta+\epsilon({\bf k})-\epsilon({\bf k}+{\bf q})} .  
\end{eqnarray}
\end{mathletters}
The Hartree solution becomes unstable if
\begin{eqnarray}
  \label{eq:HF-instab}
  1+X_{\sigma\sigma'}({\bf q},0)&=&0 .
\end{eqnarray}
This happens first for the singlet $X_{\uparrow\downarrow}$ bubble at a
critical field $B_{c0}$. It can be understood as a separator between the
weak and intermediate coupling regimes, since iterations from the Hartree
solution get unstable beyond the stability limit,
Fig.~\ref{fig:hartree-d3}.
The critical field $B_{c0}$ coincides at zero temperature with the boundary
to the saturated ferromagnetic solution $B_{s}$ \cite{vanDongen94}. This
holds even in the weak-coupling regime with no singularity in the fully
saturated ferromagnetic state. For if we get below the line of instability
of the fully saturated ferromagnet, defined by $B_s=w-U/2$, the left-hand
side of (\ref{eq:HF-instab}) becomes negative and logarithmically diverges
with a nonvanishing weight $\rho(B+Um/2)$, where $\rho$ is the density of
states and $m<1$.

\begin{figure}
  \epsfig{figure=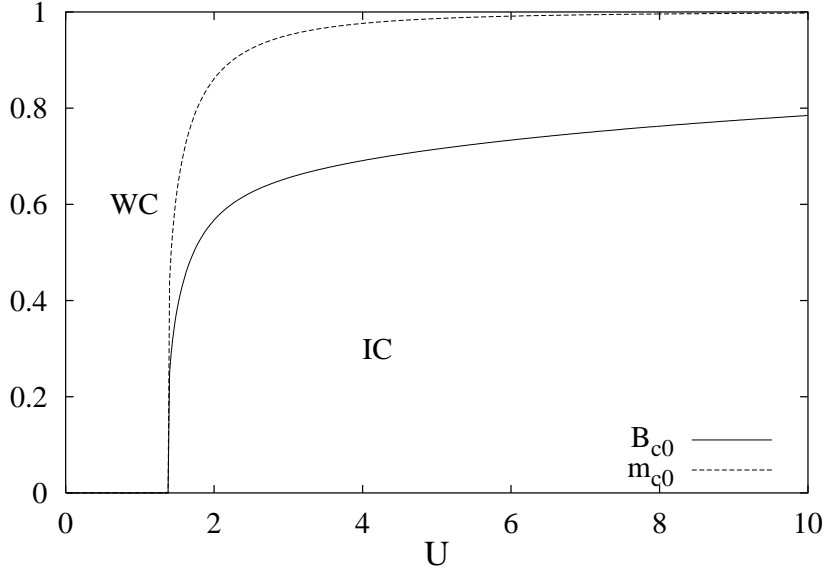,height=12cm, angle=-90}
\caption{\label{fig:hartree-imp} Instability of the Hartree approximation
  for the half-filled SIAM in an external magnetic field at zero
  temperature. The dashed line is the local magnetization along the instability
  line $m_{c0}$.}
\end{figure}

We now perform the same analysis for the SIAM in a Bethe lattice and the
Hubbard model in $d=\infty$ dimensions where only frequency variables are
relevant and no long-range order can arise.  However, the strong-coupling
phase may get insulating. We choose the following value of the
hybridization $V=w/\sqrt{2}$ with the energy unit set by the energy
half-bandwidth $w=1$.  The electron-hole bubble of an impurity embedded in
a lattice with a finite bandwidth has two contributions, band and impurity
ones. They are at zero temperature
\begin{mathletters}
\begin{eqnarray}
  \label{eq:HF-SIAM-gamma-b}
  \Gamma(0)^b_{\uparrow\downarrow}&=&-U\int_{-1}^{-1+2B}\frac{dx}{\pi}\
  \frac{\sqrt{1-x^2}}{U^2m^2/4+1-x^2}\ \frac 1{Um/2+\sqrt{(x-2B)^2-1}}
  \nonumber\\  
  &&-\frac{U^2m}2\int_{-1+2B}^B\frac{dx}{\pi}\ \frac{\sqrt{1-x^2}+\sqrt{1-
      (x-2B)^2}}{\left[U^2m^2/4+1-x^2\right]\left[U^2m^2/4+1-(x-2B)^2
    \right]} , 
\end{eqnarray}
\begin{eqnarray}
  \label{eq:HF-SIAM-gamma-I}
  \Gamma(0)^I_{\uparrow\downarrow}&=&-\frac{U^2m}{2\sqrt{1+U^2m^2/4}}\
  \frac 1{Um/2+\sqrt{\left(2B+\sqrt{1+U^2m^2/4}\right)^2-1}} 
\end{eqnarray}
\end{mathletters}
where the local magnetization is
\begin{eqnarray}
  \label{eq:HF-SIAM-m}
  m&=&\frac{Um}{2\sqrt{1+U^2m^2/4}}+\int_{-B}^B\frac{dx}\pi
  \frac{\sqrt{1-x^2}}{U^2m^2/4+1-x^2} . 
\end{eqnarray}
Evaluating the instability condition $\Gamma^b(0)+\Gamma^I(0)=-1$ we obtain
a critical field beyond which the Hartree approximation is no longer
stable.  It is plotted at zero temperature in Fig.~\ref{fig:hartree-imp}.
The boundary field at which the fully polarized solution ($m=1$) breaks
down is independent of the interaction strength, $B_s=w$, \cite{Janis99b}.
The Hartree solution becomes exact in the fully spin-polarized state. At
$B=0$, the instability of the Hartree solution with respect to a spin flip,
$U_{c0}$, lies below the interaction $U_F=2w$ at which the solution breaks
into a spin polarized state with $m\neq0$.

\begin{figure}
  \epsfig{figure=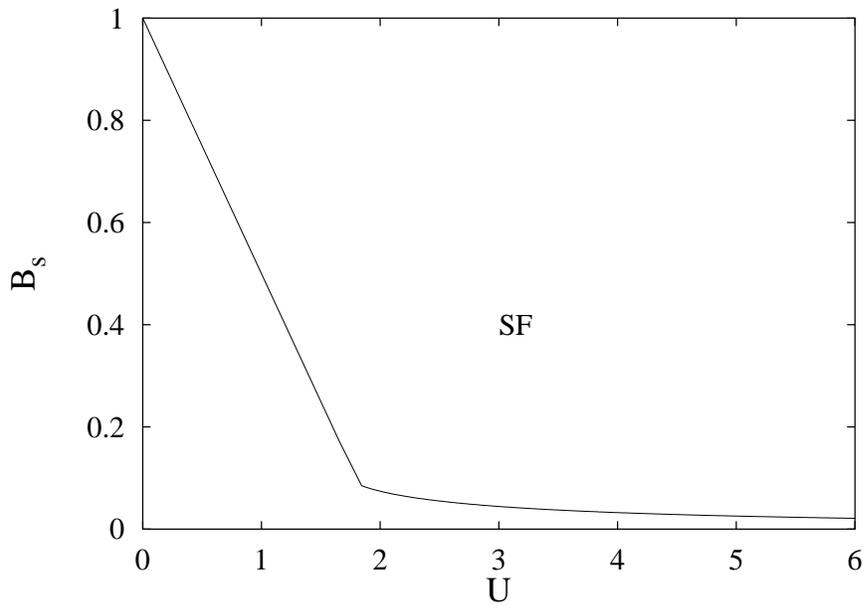,height=12cm, angle=-90}
\caption{\label{fig:bcrit-mf} Boundary field $B_s$ for the fully polarized
  solution of the Hubbard model at $T=0$ on a $d=\infty$ Bethe lattice with
  local two-particle functions. The boundary field for single impurity is
  $B_s=w=1$.}
\end{figure}

\begin{figure}
  \epsfig{figure=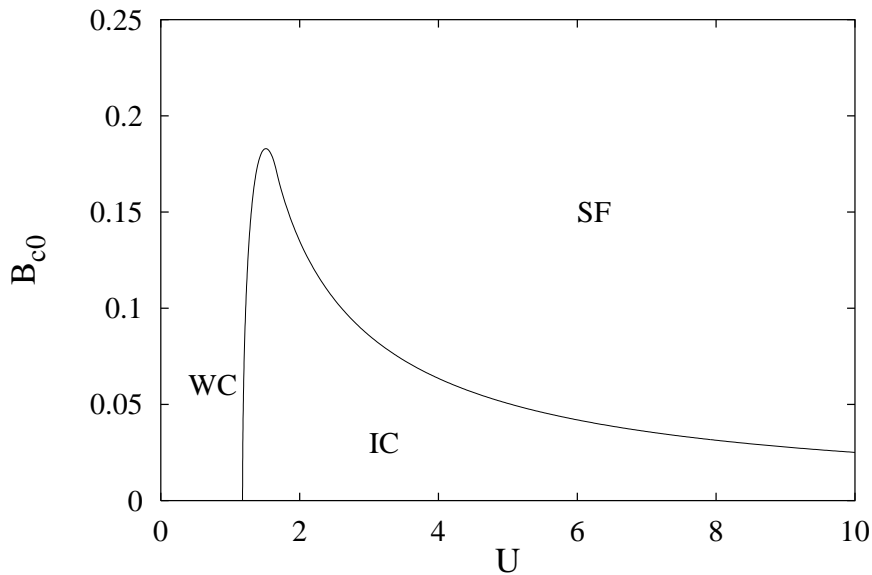,height=12cm, angle=-90}
\caption{\label{fig:hartree-mf} Instability of the Hartree approximation
  for the half-filled mean-field solution of the Hubbard model in an
  external magnetic field at zero temperature. The upper region (SF) is a
  fully polarized state ($m=1$).}
\end{figure}

A mean-field solution with only frequencies (i.e. neglecting any long-range
order) behaves differently. The singlet electron-hole bubble for a Bethe
lattice reads
\begin{mathletters}
  \begin{eqnarray}
    \label{eq:HF-MF-gamma}
    \Gamma(0)_{\uparrow\downarrow}&=&-\frac{8U}\pi\left\{\frac 23\left[1
        -\left(B+\frac U2m\right)^2\right]^{3/2}+\frac \pi2\left(B+\frac
        U2m\right)m\right\}\nonumber\\[2pt]
   && +4U\int_{-1}^{-1+2B+Um}\frac{dx}\pi\sqrt{1-x^2}\sqrt{(2B+Um-x)^2-1}    
  \end{eqnarray}
  with the magnetization defined by an equation
  \begin{eqnarray}
    \label{eq:HF-MF-m}
   m&=&\frac 2\pi\left[\arcsin\left(B+\frac U2 m\right)+\left(B+\frac U2 m
    \right)\sqrt{1-\left(B+\frac U2 m\right)^2}\ \right] .  
  \end{eqnarray}
\end{mathletters}
The boundary field for the fully polarized solution is now a function of
the interaction strength alike in finite-dimensional models,
Fig.~\ref{fig:bcrit-mf}.  The instability of the Hartree approximation
outside the fully polarized solution is plotted in
Fig.~\ref{fig:hartree-mf}.  Unlike finite-dimensional cases, the
weak-coupling regime is robust at zero temperature, since we suppressed any
possible long-range order.  At $B=0$, the instability of the Hartree
solution appears at $U_{c0}=3\pi w/8<U_F=\pi w/2$.  It lies deep below the
expected metal-insulator transition at $U_c\approx 3w$ \cite{Georges96}. It
means that the Hartree approximation for effective impurity models gets
unstable and unreliable before a spin-polarized or insulating solution may
appear. We must go to more sophisticated approximations if $U>w$ and we
want to approach the expected metal-insulator transition.

\section{Second order and single-channel approximations}
\label{sec:second-order}

An instability in the Hartree approximation in the electron-hole channel,
i.e. for the electron-hole singlet bubble, indicates either the existence
of a symmetry breaking with anomalous functions (in an external magnetic
field) or an instability with respect to dynamical fluctuations. They can
smear out the static Hartree instability and keep the symmetric solution
stable. We now check stability of simple extensions of the Hartree
mean-field theory.

\subsection{Self-consistent second order}
\label{sec:SCSOPT}

First step including dynamical fluctuations beyond Hartree is second-order
perturbation theory. We obtain this approximation if we replace the full
vertex function $\Gamma$ in (\ref{eq:sigma-2P}) by the bare coupling
constant $U$. Since the right-hand side of (\ref{eq:sigma-2P}) contains no
infinite series, the only stability criterion for this approximation is
that with (\ref{eq:2IP-diff}). Alike the Hartree approximation we expect
that the instability arises first in the electron-hole channel with  spin
singlet bubbles
\begin{eqnarray}
  \label{eq:SOPT-vertex}
  \left[\frac{\delta\Sigma}{\delta G}GG\right]^{eh}_{\sigma-\sigma}[q](k,k') 
  &=& U(1-X_{\sigma-\sigma}(q))G_{\sigma}(k')G_{-\sigma}(k'+q)
\end{eqnarray}
where $X_{\sigma\sigma'}(q)$ was defined in (\ref{eq:2P-bubble}). We have
to find the lowest real eigenvalue of the above expression seen as a matrix
in $k$ and $k'$. Since the matrix does not explicitly depend on $k$, the
lowest eigenvalue is just a sum over the other variable $k'$. We then have
a stability criterion
\begin{eqnarray}
  \label{eq:SOPT-stability}
 X_{\sigma-\sigma}({\bf q},0)\left(1 -X_{\sigma-\sigma}({\bf q},0)\right)
 &\ge& -1 . 
\end{eqnarray}
The linear term on the left-hand side origins from the Hartree
approximation. The other term from second-order theory formally enhances
the tendency to instability of the Hartree solution, since
$X_{\sigma-\sigma}({\bf q},0)<0$. However, the self-consistence
renormalizing the one-electron propagator may push the instability to
slightly higher values of the interaction strength.

\subsection{Iterated perturbation theory}
\label{sec:IPT}

Iterated perturbation theory is based on the impurity second-order
perturbation theory \cite{Georges92,Georges96}. It seems to be up to now
the only simple diagrammatic theory capable to describe both the metallic
(Fermi liquid) as well as the insulating (atomic limit) phases. This
mean-field theory is partially self-consistent with frequencies as
dynamical variables. A correction to the Hartree self-energy
$\Delta\Sigma_\sigma$ within IPT reads
\begin{eqnarray}
  \label{eq:IPT-self-energy}
  \Delta\Sigma_\sigma(i\omega_n)&=&-\frac U{\beta }\sum_m {\cal G}_\sigma(i
  \omega_n+i\nu_m){\cal X}_{-\sigma-\sigma}(i\nu_m) 
\end{eqnarray}
where
\begin{eqnarray}
  \label{eq:G-cal}
  {\cal G}_\sigma(z)&=&\frac{G_\sigma(z)}{1+G_\sigma(z)\Sigma_\sigma(z)}
\end{eqnarray}
and ${\cal X}$ is a two-particle bubble defined as in (\ref{eq:2P-bubble})
but with the frequency-dependent (local) propagator ${\cal G}$. We define a
new two-particle quantity
\begin{eqnarray}
  \label{eq:IPT-product}
  \langle{\cal X}_{\sigma\sigma'}(i\nu_m)\rangle&=& \frac U\beta\sum_n
    \frac 1{1+G_\sigma(i\omega_n)\Sigma_\sigma(i\omega_n)}\ \frac
    1{1+G_{\sigma'}(i\omega_n+i\nu_m)\Sigma_{\sigma'} 
    (i\omega_n+i\nu_m)}\nonumber \\ &&\hspace*{-40pt}\times \int d
    \epsilon\rho(\epsilon) 
    \frac  1{i\omega_n+\mu_\sigma-\Sigma_\sigma(i\omega_n)-\epsilon}\ \frac
    1{i\omega_n+i\nu_m+\mu_{\sigma'} -\Sigma_{\sigma'}(i\omega_n+i\nu_m)
    -\epsilon} 
\end{eqnarray}
needed for the stability criterion of IPT solutions. We easily find that
the IPT solution is stable w.r.t spin flips if
\begin{eqnarray}
  \label{eq:IPT-stability}
  X_{\sigma-\sigma}(0)-{\cal X}_{\sigma-\sigma}(0)\left[\langle
    {\cal X}_{\sigma-\sigma}(0)\rangle-{\cal X}_{\sigma-\sigma}(0)
  \right]&\ge&-1 .   
\end{eqnarray}
We see that IPT is generally more stable (to higher values of the coupling
constant) than the fully self-consistent perturbation theory from the
preceding subsection. It is due to subtraction in the parentheses. This
difference stems from a specific dependence of the one-particle propagator
${\cal G}$ on the self-energy.  The closer to the impurity case we are
(narrow energy band) the smaller the dynamical contribution to the
instability of the IPT solution is. In the extreme case of the impurity
model, there is no contribution to the instability from second order, i.e.
beyond Hartree.  Because of the Hartree term there is always a critical
interaction at which the IPT solution gets unstable in both the single
impurity and the mean-field.  This instability, as we already know from the
Hartree solution, lies well below the expected Mott-Hubbard metal-insulator
transition. It is hence inappropriate to extrapolate a metallic IPT
solution to the critical region near the Mott-Hubbard transition and
beyond. It seems to be no way of matching consistently the strong-coupling
(insulating) and the weak-coupling (metallic) IPT solutions.

IPT from mean-field models amounts to a non-self-consistent expansion on
finite dimensional lattices.  A non-self-consistent expansion is derived
within the general scheme of Section~\ref{sec:stability} if the
renormalized one-electron propagator is replaced by
$G^{(0)-1}=G^{-1}+\Sigma[G]$ where the self-energy is treated as a
functional of the full one-electron propagator. All orders of the
non-self-consistent expansion get unstable at the same point where the
Hartree approximation does. A non-self-consistent expansion does not
improve upon stability of the Hartree solution.

\subsection{Single-channel approximations}
\label{sec:SCh}

Dynamical approximations cannot push the instability of the Hartree
solution to much higher interaction strengths unless they contain
infinite-many two-particle bubbles.  Simplest theories with geometric
series of two-particle bubbles are single-channel approximations or their
extension, the so-called FLEX approximation \cite{Bickers89}. I.~e. we
explicitly sum multiple two-particle scatterings on the bare interaction
within a single two-particle channel. We derive these approximations within
the general scheme of Section~\ref{sec:stability} if we choose
$\Lambda^\alpha=U,\ \Lambda^{\alpha'}=0,\ \alpha'\neq\alpha$
\cite{Janis98a}. Since we expect that the instability arises in the
electron-hole channel, we explicitly consider only a renormalized RPA
(electron-hole channel).  A correction to the Hartree self-energy for this
approximation is
\begin{eqnarray}
  \label{eq:SCh-self-energy}
  \Delta\Sigma_\sigma(k)&=&-\frac U{\beta N}\sum_q G_{-\sigma}(k+q)\
  \frac{X_{\sigma-\sigma}(q)}{1+X_{\sigma-\sigma}(q)} .
\end{eqnarray}
The full vertex function $\Gamma$ from the Schwinger-Dyson equation
(\ref{eq:sigma-2P}) already contains a geometric series of two-particle
bubbles, so we have two nontrivial stability criteria. The one derived from
the equation of motion in this approximation reads
\begin{eqnarray}
  \label{eq:SCh-stability1}
  X_{\sigma-\sigma}({\bf q},0)&\ge&-1 .
\end{eqnarray}
This is the stability condition from the Hartree approximation with a
renormalized one-electron propagator with the self-energy from
(\ref{eq:SCh-self-energy}).  Condition (\ref{eq:SCh-stability1}) is
fulfilled for weakly and moderately correlated systems significantly beyond
the instability of the Hartree solution. In the impurity or mean-field
cases it is even satisfied up to infinite interaction strength
\cite{Hamann69}. A solution fulfilling (\ref{eq:SCh-stability1}) is
internally consistent and possesses the necessary analytic properties of
the self-energy.  It, however, does not guarantee stability of the
approximation with respect to external perturbations and to fluctuations
beyond this solution. For that purpose we use definition
(\ref{eq:2IP-diff}) for the irreducible functions. With the generic
notation (\ref{eq:channels}) we obtain a new vertex function in the
electron-hole channel
\begin{eqnarray}
  \label{eq:SCh-vertex}
  \left[\frac{\delta\Sigma}{\delta G}\right]^{eh}_{\sigma-\sigma}[q](k,k') 
  &=& \frac U{1+X_{\sigma-\sigma}(k'-k)}\nonumber\\&& -\frac{U^2}{\beta N}
  \sum_{k''}\frac{G_{-\sigma}(k+k'+q-k'')}{1+X_{\sigma-\sigma}(k'+q-k'')}\
  \frac{G_{\sigma}(k'')}{1+X_{\sigma-\sigma}(k'-k'')}  .
\end{eqnarray}
This vertex function enters the irreducible functions in the other
channels, the interaction and the electron-electron ones. We identify in
each channel the conserving bosonic variable according to rules
(\ref{eq:channels}) and obtain a matrix in $k,k'$ that is to be
diagonalized. Unlike second-order we cannot find the lowest real eigenvalue
in closed form. We can at least qualitatively assess the lowest eigenvalue
in the asymptotic limit of condensation of electron-hole pairs, where the
electron-hole two-particle vertex function diverges. First, the
electron-electron channel, at least at half filling, is irrelevant and we
resort to the electron-hole one. Next, we neglect the convolution of the
singular vertices with regular electron-hole bubbles assuming that the
integrals smear out the singularities. We hence keep only the first term on
the right-hand side of (\ref{eq:SCh-vertex}) as a leading divergent
contribution in the asymptotic limit of the critical point.  Even then we
still have a full $k,k'$ matrix to diagonalize. We further resort to zero
temperature with no difference between fermionic and bosonic variables and
assume that the lowest eigenvalue can be approximated as in second-order
theory. I.~e. we treat the matrix as if it were $k$-independent. This
assumption does not hold strictly but only approximately if the influence
of the states away from the Fermi surface is not too big.  Within this
approximation we obtain a new approximate thermodynamic stability criterion
in the interaction channel
\begin{eqnarray}
  \label{eq:SCh-stability2}
  \frac U{\beta N}\sum_{q'} \frac{G_{-\sigma}(({\bf k}_0,0)+q')G_{-\sigma}
    (({\bf k}_0,0)+q'+({\bf q},0))}{1+X_{\sigma-\sigma}(q')}&\ge&-1 
\end{eqnarray}
where ${\bf k}_0$ is a suitable vector from the Fermi surface minimizing
the expression on the left-hand side.

We now have two criteria that both must be fulfilled in order to have a
stable phase with correct analytic properties of the Green functions. The
former criterion, (\ref{eq:SCh-stability1}), from the equation of motion
guarantees internal consistency and analyticity of the approximation.  The
latter one, (\ref{eq:SCh-stability2}), when fulfilled, implies stability of
the approximation with respect to fluctuations beyond the renormalized RPA.
When broken, scatterings in the neglected interaction channel become
relevant and must self-consistently be interwoven with the electron-hole
ones. Simple additions of channels as in the FLEX approximation do not
introduce a self-consistence in the two-particle functions and do not
improve upon the stability of the single-channel approximations. In low
dimensions ($d=2,1$), the left-hand side of (\ref{eq:SCh-stability2}) may
even diverge and the single-channel approximations become qualitatively
wrong, since the full solution can produce only integrable singularities.
Hence FLEX-type approximations are unsuitable for the description of the
weak-to-strong-coupling transition.

\section{Parquet approximation}
\label{sec:parquet}

The approximations we have hitherto dealt with were either derived by
simplifying the full two-particle vertex function $\Gamma$ in the
Schwinger-Dyson equation or some of the Bethe-Salpeter equations were
disregarded.  The various two-particle functions were hence treated
inequivalently and some divergences went lost.  The simplest approximation
consistent with all the Bethe-Salpeter equations (\ref{eq:Bethe-Salpeter})
is the parquet approximation with the unrenormalized completely irreducible
vertex $I[U;G,\Gamma]=U$. We formulate explicitly the stability criteria
for this approximation and show how the singularities from the
Bethe-Salpeter equations are controlled.

To control the singularities that may arise in two-particle
functions we diagonalize the Bethe-Salpeter convolutive equations
(\ref{eq:Bethe-Salpeter}).  We, however, can diagonalize the Bethe-Salpeter
equations only separately in each channel and cannot express the solution
in closed algebraic form. Diagonalization in the chosen two-particle
channel means diagonalization of the appropriate matrix from
(\ref{eq:channels}). To formalize this procedure we represent the one- and
two-particle functions as operators on a Hilbert space of two-particle
(bare) states. Such a Hilbert space can be spanned on the basis vectors
$|k,k'>$ parameterized by two one-particle four-momenta. The two-particle
vertex functions can be represented as
\begin{eqnarray}
  \label{eq:parquet-vertex}
  \widehat{X}_{\sigma\sigma'}&=&\sum_{kk'\bar{k}\bar{k}'}
  \delta(k+k'-\bar{k}-\bar{k}')X_{\sigma\sigma'}(k,k';\bar{k},
  \bar{k}') |kk'\rangle\langle \bar{k}'\bar{k}|  .
\end{eqnarray}
We now switch from this channel-independent representation to
channel-dependent diagonal representations. We use $q$ for the conserved
four-momenta and $Q$ for the eigenvalues of the corresponding matrices in
the two-particle channels from (\ref{eq:channels}). 

We need diagonal representations for the $\alpha$-reducible two-particle
functions
\begin{eqnarray}
  \label{eq:parquet-2PR}
  \widehat{{\cal K}}^\alpha&=&\sum_{q,Q}{\cal K}^\alpha(q,Q)\ |Qq,\alpha
  \rangle\langle\alpha,qQ| .
\end{eqnarray}
Note that the eigenvectors $|Qq,\alpha>$ are expressed as a combination of
the vectors $|k,k'>$ where $Q$ is a combination of $k$'s.  The vector $q$
fixes either difference (electron-hole and interaction channels) or sum
(electron-electron channel) of $k',k$. All $|k,k'>$ and $|Qq,\alpha>$ form
complete orthogonal sets. For the sake of simplicity we suppressed
normalization factors in the sums in (\ref{eq:parquet-vertex}) and
(\ref{eq:parquet-2PR}).  Representation (\ref{eq:parquet-2PR}) makes the
Bethe-Salpeter equations algebraic and we can write their formal solutions
as
\begin{mathletters}\label{eq:parquet-solution}
\begin{eqnarray}
  \label{eq:parquet-diagonal}
  {\cal K}^\alpha(q,Q)&=&-\frac{\langle\alpha,qQ|\ \widehat{X}^\alpha\ |Qq,
    \alpha\rangle}{1+ \langle\alpha,qQ|\ \widehat{Y}^\alpha\ |Qq,\alpha
    \rangle} 
\end{eqnarray}
where $\widehat{X}$ and $\widehat{Y}$ are two-particle operators. They have
the following explicit representation
\begin{eqnarray}
  \label{eq:parquet-X}
  \langle\alpha,qQ|\ \widehat{X}^\alpha\ |Qq,\alpha\rangle&=& \langle
  \alpha,qQ|\widehat{U}\widehat{G}^{(2)}\widehat{U}|Qq,\alpha
  \rangle+\sum_{\alpha'\neq\alpha}\sum_{q',Q'}{\cal K}^{\alpha'}(q',Q')
  \nonumber\\ &&\hspace*{-110pt}
  \times\left[\langle\alpha,qQ|\widehat{U}\widehat{G}^{(2)}|Q'q',
    \alpha'\rangle \langle\alpha',q'Q'|Qq,\alpha\rangle+  \langle\alpha,qQ
    |Q'q',\alpha'\rangle \langle\alpha',q'Q'|\widehat{G}^{(2)}
    \widehat{U}|Qq,\alpha\rangle\right] \nonumber\\ &&\hspace*{-110pt}
  +\sum_{\alpha'\alpha''\neq\alpha}\ \sum_{q'q'',Q'Q''}
  {\cal K}^{\alpha'}(q',Q') {\cal K}^{\alpha''}(q'',Q'')
  \langle\alpha,qQ| Q'q',\alpha'\rangle\langle\alpha',q'Q'|\widehat{G}
  ^{(2)}|Q''q'',\alpha''\rangle\nonumber\\  
  &&\times\langle\alpha'',q''Q''| Qq,\alpha\rangle ,
\end{eqnarray}
and
\begin{eqnarray}
  \label{eq:parquet-Y}
   \langle\alpha,qQ|\ \widehat{Y}^\alpha\ |Qq,\alpha\rangle&=&\langle
   \alpha,qQ|\widehat{U}\widehat{G}^{(2)}|Qq,\alpha
   \rangle+\sum_{\alpha'\neq\alpha}\sum_{q',Q'}{\cal K}^{\alpha'}(q',Q') 
   \nonumber\\ &&
   \times\langle\alpha,qQ|\widehat{U}\widehat{G}^{(2)}|Q'q',\alpha'
   \rangle\langle\alpha',q'Q'|\widehat{G}^{(2)}| Qq,\alpha\rangle .  
\end{eqnarray}
\end{mathletters}
We used abbreviations for the free two-particle propagation
\begin{mathletters}\label{eq:parquet-aux}
\begin{eqnarray}
   \label{eq:parquet-G2}
   \widehat{G}^{(2)}_{\sigma\sigma'}&=&\sum_{kk'}G_\sigma(k)G_{\sigma'}(k') 
   |kk'\rangle\langle k'k|
 \end{eqnarray}
 and for the operator of the bare vertex
\begin{eqnarray}
  \label{eq:parquet-U}
  \widehat{U}_{\sigma\sigma'}&=&U\delta_{\sigma',-\sigma}\sum_{kk'\bar{k}
  \bar{k}'} \delta(k+k'-\bar{k}-\bar{k}')\ |kk'\rangle\langle \bar{k}'
  \bar{k}| .
\end{eqnarray}
\end{mathletters}
The parquet approximation has become a set of closed equations for ${\cal
  K}^\alpha(q,Q)$ for the given one-electron propagators. The most
difficult task in solving parquet equations (\ref{eq:parquet-solution}),
(\ref{eq:parquet-aux}) is the diagonalization in the separate channels,
i.~e.  finding the bases $|Qq,\alpha>$ represented in the bare states
$|kk'>$. They are determined from (\ref{eq:channels}) and
(\ref{eq:2IP-def}) so that the parquet equations remain implicit and must
be iterated.  However, parquet equations (\ref{eq:parquet-solution}) allow
us to control the behavior of the denominators, i.~e. the appearance of
singularities.

To close the parquet approximation we have to rewrite the Schwinger-Dyson
equation with the diagonal form of the vertex function. It reads
\begin{eqnarray}
  \label{eq:parquet-sigma}
  \Delta\Sigma_{\sigma}(k)&=&-\sum_{k'}G_{-\sigma}(k')\left[\langle k'k|
    \widehat{U}\widehat{G}^{(2)}_{\sigma-\sigma}\widehat{U}|kk'\rangle
    +\sum_{\alpha}\sum_{q,Q}{\cal K}^{\alpha}_{\sigma-\sigma}(q,Q)
  \right.\nonumber\\ &&\hspace*{80pt} \times
  \langle k'k|Qq,\alpha\rangle\langle\alpha,qQ|\widehat{U}\widehat{G}^{(2)} 
  _{\sigma-\sigma}|kk'\rangle\Bigg] .   
\end{eqnarray}
The parquet approximation is completely determined by numbers
$\Sigma_\sigma(k)$, ${\cal K}^\alpha_{\sigma\sigma'}(q,Q)$ and
$<\alpha,qQ|kk'>$.  The internal stability criteria for the parquet
approximation in the diagonal representation are
\begin{eqnarray}
  \label{eq:parquet-stability}
  1+ \langle\alpha,({\bf q},0)({\bf Q},0)|\ \widehat{Y}^\alpha\ |({\bf Q},0)
  ({\bf q},0),\alpha\rangle&\ge&0 .
\end{eqnarray}
We see that all the two-particle functions can contain only integrable
singularities and that if a singularity arises in one channel it cannot
cause a havoc in the other channels, since the integral kernels are
integrable functions.

There are also stability criteria for the parquet approximation with
respect to external perturbations beyond this approximate equilibrium
state. Since possible instabilities must be integrable we do not expect a
qualitatively different behavior. A qualitative change could be caused by
new bound states with more than two particles that could not be decomposed
into simpler two-particle states. It is physically plausible to assume
nonexistence of such states, at least all the existing approaches do that.

It is a tremendous task to solve the complete parquet approximation. It can
be completed only in a very limited special situations \cite{Bickers98}.
However, the general scheme how to solve the parquet equations helps us to
identify the principal features any simplification should not loose. It is
a two-particle self-consistence allowing only for integrable singularities
and an access to the diagonal or algebraic form of the solution of the
Bethe-Salpeter equations enabling to control possible singularities.

\section{Simplifications in the parquet approach}
\label{sec:simplifications}

Parquet diagrams differ significantly from the simpler single-channel
approximations with multiple two-particle scatterings. The difference
manifests itself in the effective interaction in the two-particle
scattering processes. The bare interaction remains unrenormalized in the
single-channel approximations. In the parquet equations, it is replaced by
renormalized vertex functions representing a dynamically screened
inter-particle potential.  A qualitative difference between a solution to
the full parquet equations and to the single-channel approximations becomes
evident at the two-particle criticality. The vertex functions from one or
more channels get critical and sharply peaked around the Fermi energy. The
exchange between the scattered particles becomes strongly energy dependent
and cannot be approximated by a static effective interaction as in the
single-channel theories.

The Bethe-Salpeter nonlinear integral equations are extremely complicated
with an intricate analytic structure of the solution.  Each two-particle
function entering the parquet equations contains initially three
independent (complex) variables. Neither of them can simply be neglected,
since the variables are interconnected due to a mixing of different
inequivalent channels.  Even the diagonal form of the parquet equations
with only two relevant variables remains implicit and does not allow for a
solution in closed form.  That is why a nonperturbative solution to the
parquet equations has not yet been found.  From this reason the parquet
diagrams are relatively little used in condensed matter although being in
existence more than thirty years.

The only hope that the parquet approximation may become useful for
quantitative assessments of the effects of strong electron correlations is
that one succeeds in simplifying the full parquet algebra to an
analytically manageable form. As already discussed, the parquet approach is
expected to produce qualitatively new results at intermediate and strong
coupling with critical two-particle functions where weak-coupling
approximations get unstable. In simplifying the parquet equations we will
demand asymptotic accuracy at the two-particle criticality, particularly
for condensation of electron-hole pairs leading at half filling to a
metal-insulator transition.  We hence retain only the diagrams potentially
making the effective mass of the electrons divergent at the metal-insulator
transition.

\subsection{Two-Variable Two-Particle Functions}

It was recently argued that at least at half filling only two of the three
topologically inequivalent channels are relevant in the critical region of
the metal-insulator transition. Only multiple scatterings in the
electron-hole and the interaction channels contain divergent diagrams at
the critical point \cite{Janis98a}. This simplification to two channels
itself, called dipole approximation, does not reduce the number of
variables in the two-particle functions.  It only simplifies the algebra of
the parquet approximation. To lower the number of the relevant variables we
use the fact that the singularities in the two-particle functions in the
parquet approximation must be integrable. It means that when integrating
over the variable in which the singularity arises we obtain a finite result
bounded by a constant of order one. Our approximation consists in
neglecting all finite (nondiverging) contributions at the critical point.
This must be done at the level of the diverging two-particle functions.

As a first step we reduce the number of the relevant variables in the
two-particle functions to two. This is achieved when we neglect mixing of
the reducible functions from different channels. We replace the full
two-particle function $\Gamma$ on the right-hand side of the Bethe-Salpeter
equation in the $\alpha$-channel by $U+{\cal K}^\alpha$, i.~e. by a sum of
the completely irreducible vertex and the reducible function in that
channel. Such a replacement does not change the denominator of the diagonal
representation of the solution to the Bethe-Salpeter equations and hence
the universal critical behavior of the function ${\cal K}^\alpha$ remains
unaltered. The bare interaction is a constant and consequently the solution
does not depend on the outgoing (incoming) variables. Further on, the
triplet functions can be eliminated from this approximation, since
$\Lambda_{\sigma,\sigma}$ gets irrelevant.  If we denote
$\Lambda^v_{\sigma,-\sigma}=U+{\cal K}^{eh}_{\sigma,-\sigma}$ and
$\Lambda^h_{\sigma,-\sigma}=U+{\cal K}^U_{\sigma,-\sigma}$, the parquet
equations reduce to
\begin{mathletters}\label{eq:I}
\begin{eqnarray}
  \label{eq:parq2a}
  \Lambda^v_{\sigma,-\sigma}(k;q)&=&U\nonumber\\ &&\hspace*{-10pt} -\frac
  1{\beta{\cal N}}\sum_{q'}\Lambda^h_{\sigma,-\sigma}(k;q')G_\sigma(k+q')
  G_{-\sigma} (k+q'+q) \Lambda^v_{\sigma,-\sigma}(k+q';q)\ , \\[4pt]
  \label{eq:parq2b} 
  \Lambda^h_{\sigma,-\sigma}(k;q)&=&U+\frac 1{\beta^2{\cal N}^2}\sum_{q',q''}
  \Lambda^v_{\sigma,-\sigma}(k;q')G_{-\sigma}(k+q') G_{-\sigma}(k+q'+q)
  \nonumber\\
 &&\hspace*{-10pt} \times \Lambda^v_{-\sigma,\sigma}(k+q';q''-q') G_\sigma
 (k+q'') G_\sigma (k+q''+q)\Lambda^h_{\sigma,-\sigma}(k+q'';q)\ . 
\end{eqnarray}\end{mathletters}

Because of the broken symmetry between the incoming and outgoing variables
in the two-particle functions we must be careful in simplifying the
equation for the self-energy. To be consistent we use a symmetrized formula
with the bare interaction at the incoming and outgoing momenta. We obtain
\begin{eqnarray}
  \label{eq:Sigma2}
  \Delta\Sigma_\sigma(k)&=&-\frac U{2\beta^2{\cal N}^2}\sum_{k',q} G_\sigma(k+q)
  G_{-\sigma}(k')  G_{-\sigma}(k'+q)\left[\Lambda^h_{\sigma,-\sigma}(k+q;
  -q)\right.     \nonumber\\ 
   &&\left. + \Lambda^h_{-\sigma,\sigma}(k';q)+ \Lambda^v_{\sigma,-\sigma}(k+q;k'
  -k)+\Lambda^v_{-\sigma,\sigma}(k';k-k') -2U\right]\ .
\end{eqnarray}

The reduced two-channel parquet approximation with two-variable
two-particle functions reproduces in leading order the critical behavior of
the parquet approximation (with two channels). The universal quantities
derived from the divergent functions of the parquet equations are
reproduced by equations (\ref{eq:I})-(\ref{eq:Sigma2}). Only nonuniversal
quantities such as the critical interaction do not come out in this
simplification as in the full parquet theory.

We achieved a simplification of the full parquet algebra without loosing
leading critical behavior of the solution, but the resulting equations
retain a nonlinear convolutive character. It is still too elaborate to
reach nonperturbative solutions or to diagonalize them. We must find a
further reduction of complexity of the parquet equations to make them
useful for an affordable quantitative analysis.

\subsection{One-Variable Two-Particle Functions}

It is the bosonic variable $q$ in the two-particle functions
$\Lambda^h,\Lambda^v$ that is relevant at the critical point and in which
the singularity arises. The fermionic variable $k$ in both functions
``labels'' the eigenvalues of the integral kernel in equations
(\ref{eq:parq2a}), (\ref{eq:parq2b}). The minimal eigenvalue governs the
critical behavior at low temperatures. If we are interested only in the
asymptotic critical behavior we can approximate the actual minimal
eigenvalue by a value at the lowest-lying fermionic four-momentum. By such
an approximation we neglect mixing of the fermionic four-momenta in the
Bethe-Salpeter equations and $k$-dependence in the two-particle functions.
This approach corresponds to a low-energy expansion used by Hamann
\cite{Hamann69} by assessing the strong-coupling limit of the renormalized
RPA of Suhl in the single-impurity problem. We assume that the approximate
minimal eigenvalue dominates and qualitatively determines the relevant
physics of the critical point.

The suggested simplification is possible only at zero temperature, where
the difference between the fermionic and bosonic Matsubara frequencies
vanishes. We hence put $T=0$, which is the most interesting case for the
metal-insulator transition. Using the above ansatz, neglecting
$k$-dependence in $\Lambda^h,\Lambda^v$, and putting $k=({\bf k}_0,0)$ in
the one-electron propagators we reduce the parquet equations to a
manageable form. We introduce new functions
\begin{mathletters}\label{eq:vertex}
\begin{eqnarray}
  \label{eq:Gam}
  \Gamma_{\sigma}(q)&=&\frac 1{\beta{\cal N}}\sum_{q'}
  \Lambda^h_{\sigma,-\sigma}(q') G_{\sigma}(({\bf k}_0,0)+q')G_{-\sigma}
  (({\bf k}_0,0)+q'+q)\\  \label{eq:K}
  {\cal K}_{\sigma}(q)&=& \frac 1{\beta{\cal N}}\sum_{q'}
  \Lambda^v_{-\sigma,\sigma}(q') G_{\sigma}(({\bf k}_0,0)+q')G_{\sigma}
  (({\bf k}_0,0)+q'+q)\ , 
\end{eqnarray}\end{mathletters}
with the aid of which we obtain for the two-particle vertex functions
\begin{mathletters}
\begin{eqnarray}
  \label{eq:Iv}
  \Lambda^v_{\sigma,-\sigma}(q)&=&\frac U{1+\Gamma_{\sigma}(q)}\ ,\\[4pt]
  \label{eq:Ih} \Lambda^h_{\sigma,-\sigma}(q)&=& \frac U{1-{\cal K}
    _{-\sigma}(q){\cal K}_{\sigma}(q)} \ .
\end{eqnarray}\end{mathletters}
The introduced fermionic momentum $({\bf k}_0,0)$ from the Fermi surface is
an adjustable parameter chosen to maximize the tendency to instability of
this approximation.

The above equations have the complexity of the single-channel
approximations, but contain renormalized vertex functions in the
two-particle scatterings.  It is always the renormalized vertex function
that enters the denominator of equations (\ref{eq:I}) via the functions
$\Gamma$ and ${\cal K}$ and the critical behavior is properly renormalized.

It is easy to derive the corresponding equation for the self-energy. It
reads
\begin{eqnarray}
  \label{eq:sigma1}
  \Delta\Sigma_\sigma(k)&=&-\frac 1{2\beta{\cal N}}\sum_q\left[G_\sigma(k+q)
    X_{-\sigma,-\sigma}(q) \left(\Lambda^h_{-\sigma,\sigma}(q)+\Lambda^h
    _{\sigma,-\sigma}(-q)-U\right)\right.\nonumber\\
   &&\hspace*{20pt}\left. +G_{-\sigma}(k+q)X_{\sigma,-\sigma}(q)\left(
    \Lambda^v_{\sigma,-\sigma}(q)+I^v_{-\sigma,\sigma}(-q) -U\right)\right]
\end{eqnarray}
where $X_{\sigma,\sigma'}$ are two-particle bubbles with the renormalized
one-electron propagators.
 
Equations (\ref{eq:vertex})-(\ref{eq:sigma1}) replace the original parquet
approximation. They fully determine the generating two- and one-particle
functions. All thermodynamic and spectral quantities can be calculated from
them. Since we used a number of approximate steps in the derivation of the
reduced parquet algebra, any new quantity must first be formulated within
the full parquet approximation.  The Luttinger-Ward functional generating
all thermodynamic quantities in the parquet approximation and
simplifications thereof is constructed in the Appendix.  Simplifications
are introduced in the exact formulas via the generating one- and
two-particle functions.

Having a simplified and manageable set of parquet-type equations we can
finally formulate internal stability criteria for them. They are derived
from the general ones and resemble the stability criteria from the
single-channel approximations. They read
\begin{eqnarray}
  \label{eq:simplified-stability}
  1+{\cal K}_{\sigma}({\bf q},0)\ge 0 &,\hspace{10mm}& 1+\Gamma_{\sigma}(
{\bf q},0)\ge 0 .
\end{eqnarray}
Unlike the single-channel approximations, stability criteria
(\ref{eq:simplified-stability}) contain renormalized dynamical two-particle
vertex functions that are determined self-consistently from the simplified
parquet equations. The functions $\Gamma_\sigma$ and ${\cal K}_{\sigma}$
depend implicitly on the momentum ${\bf k}_0$ from the Fermi surface emerging
due to consistency of our simplification.  The vector ${\bf k}_0$ is chosen
so that to minimize the left-hand side of (\ref{eq:simplified-stability})
The vector then determines the character of the symmetry breaking ($s$ vs.
$d$ wave etc.) first appearing in the system.

\section{Dynamical mean-field approximation and metal-insulator transition}
\label{sec:DMFT}

We apply the dipole (simplified parquet) approximation to effective
impurity models (single impurity or dynamical mean-field) at intermediate
and strong coupling. The fluctuations in space do not contribute to the
dynamics and the four-momenta from the general formulation of the parquet
approximation collapse to frequencies.  This simplification enables one to
perform explicitly analytic continuation from the Matsubara to the real
frequencies using contour integrals. Realizing that the reduced equations
hold for $T=0$ and introducing $\zeta,z$ for the bosonic, fermionic complex
frequencies, respectively, we can write for the two-particle functions
\begin{mathletters} \label{eq:2P-cont}
\begin{equation}
  \label{eq:K-cont}
  {\cal K}_\sigma(\zeta)=-U\int_{-\infty}^0\frac{d\omega}\pi
  \left[G_\sigma(\omega+\zeta)\mbox{Im}\frac{G_\sigma(\omega_+)}
    {1+\Gamma_{-\sigma}(\omega_+)} + \frac{G_\sigma(\omega-\zeta)}
    {1+\Gamma_{-\sigma}(\omega-\zeta)}\mbox{Im}G_\sigma(\omega_+)\right], 
\end{equation} 
\begin{eqnarray}
  \label{eq:Gam-cont}
  \Gamma_\sigma(\zeta)&=&-U\int_{-\infty}^0\frac{d\omega}\pi
  \left[G_{-\sigma}(\omega+\zeta)\mbox{Im}\frac{G_\sigma(\omega_+)}
    {1-{\cal K}_\sigma(\omega_+){\cal K}_{-\sigma}(\omega_+)}
  \right. \nonumber\\[4pt]  &&\hspace*{80pt}\left.
    +\frac{G_\sigma(\omega-\zeta)}{1-{\cal K}_\sigma(\omega-\zeta){\cal 
      K}_{-\sigma}(\omega-\zeta)} \mbox{Im}G_{-\sigma}(\omega_+)\right] 
\end{eqnarray}\end{mathletters}
and analogously for the one-particle self-energy correction to the Hartree
term
\begin{eqnarray}
  \label{eq:Sigma1}
 \Delta \Sigma_\sigma(z)&=&\int_{-\infty}^0\frac{d\omega}\pi\left\{
    G_\sigma(\omega+z) \mbox{Im}\left[X_{-\sigma,-\sigma}(\omega_+)\left(
     \Lambda^h_{-\sigma,\sigma}(\omega_+) -U\right)\right]\right.\nonumber\\
  &&\hspace*{-50pt}\left. +X_{-\sigma,-\sigma} (\omega-z)\left(\Lambda
    ^h_{-\sigma,\sigma}(\omega_+) -U\right)\mbox{Im}G_\sigma(\omega_+)
  +G_{-\sigma}(\omega+z)\mbox{Im}\left[X_{\sigma,-\sigma}(\omega_+)
     \Lambda^v_{\sigma,-\sigma}(\omega_+)\right] \right.\nonumber \\
  &&\left. +X_{\sigma,-\sigma}(\omega-z)\Lambda^v_{\sigma,-\sigma}(\omega-z)
    \mbox{Im}G_{-\sigma}(\omega_+)\right\} 
\end{eqnarray}
where we denoted $\omega_+=\omega+i0^+$.

Equations (\ref{eq:2P-cont}) and (\ref{eq:Sigma1}) can be solved
numerically. We use the one-electron propagators from the Bethe lattice in
$d=\infty$ with a finite half-bandwidth $w=1$. We as well put $B=0$, since
the Mott-Hubbard metal-insulator transition is expected there and both
channels should show the same divergence at this critical point.

We have three levels of self-consistence in the parquet equations.  We have
to determine self-consistently the Hartree parameters (spin-dependent
particle densities), then the dynamical self-energy correction, and finally
the two-particle vertex functions (cf. Appendix, (\ref{eq:Omega})). It
appears that we have to perform the self-consistences sequentially from top
to bottom.  First the Hartree must be self-consistently determined, second
the self-energy correction, and finally the vertex functions.  Only in this
order we can secure convergence of the iteration scheme and analytic
properties of the converged and intermediate solutions.

\begin{figure}
  \epsfig{figure=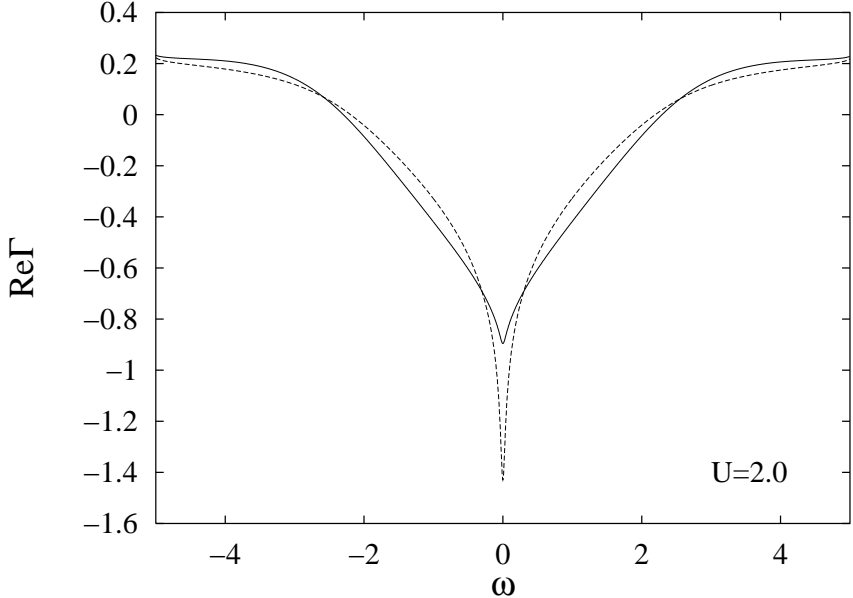,height=12cm, angle=-90}
\caption{\label{fig:FLEX-instab} Vertex function $\Gamma$ calculated within
  the FLEX (solid line) and first iteration beyond FLEX (dashed line).
  Their values at zero frequency decide about stability of the
  approximation ($\mbox{Re}\Gamma(0)>-1$).}
\end{figure}

First step in our iteration scheme for the two-particle functions reduces
to self-consistent second order.  It becomes unstable at around $U\approx
w$ and gets spin polarized at intermediate coupling.  Iterations to the FLEX
approximation with two (electron-hole and interaction) channels converge
very quickly at weak and intermediate coupling. The converged solution
never gets spin-polarized.  However, starting with $U\approx w$ the
solution becomes unstable with respect to external perturbations,
spin-flips, and condition (\ref{eq:SCh-stability2}) is violated.
Fig.~\ref{fig:FLEX-instab} shows the real part of the vertex function
$\Gamma$ for the FLEX with the first next iteration towards the parquet
solution. A stable solution demands that both functions at $\omega=0$ lie
above the value $-1$.

The iterations for the reduced dipole approximation converge at weak and
intermediate coupling for $U\le 2w$. We experienced problems with
convergence for higher values of the interaction strength \cite{note2}. Due
to logarithmic singularities we had insufficient resolution close to the
Fermi energy. We used a linear distribution of the frequency mesh points
with $\Delta\omega=0.005w$, since it is convenient for evaluating the
convolutions. We need all the energy scales to reach a solution fulfilling
necessary sum rules, but as the metal-insulator transition is approached, a
broad spectrum of energy scales is to be covered. A more sophisticated
method for dealing with the different energy scales in equations
(\ref{eq:2P-cont}), such as numerical renormalization group, has to be
applied if we want to continue the approximation closer to the
metal-insulator transition.

\begin{figure}
  \epsfig{figure=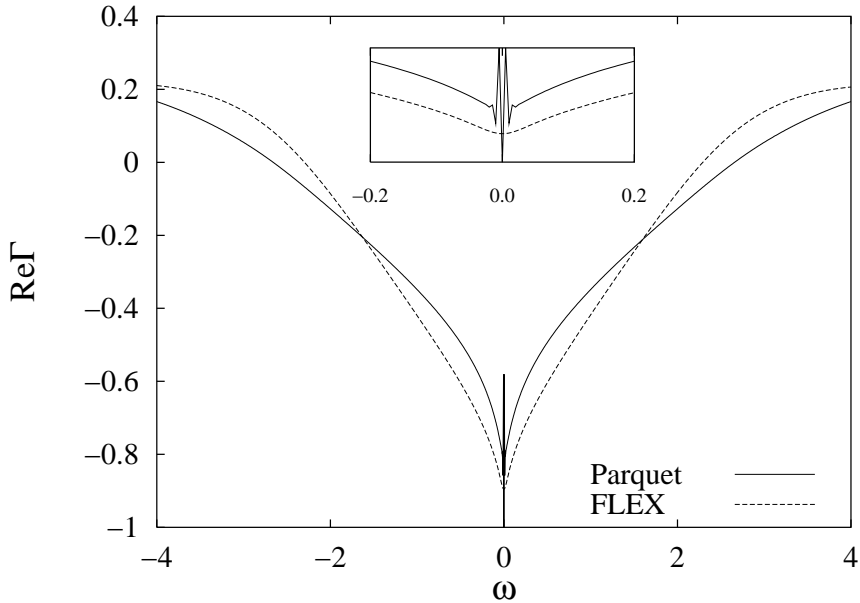,height=12cm, angle=-90}
\caption{\label{fig:FLEX-comp2}Comparison of the two-particle vertex
  calculated at $U=2w$ within the FLEX and the simplified parquet
  approximations. The inset shows the detailed structure around the Fermi
  energy.}
\end{figure}

\begin{figure}
  \epsfig{figure=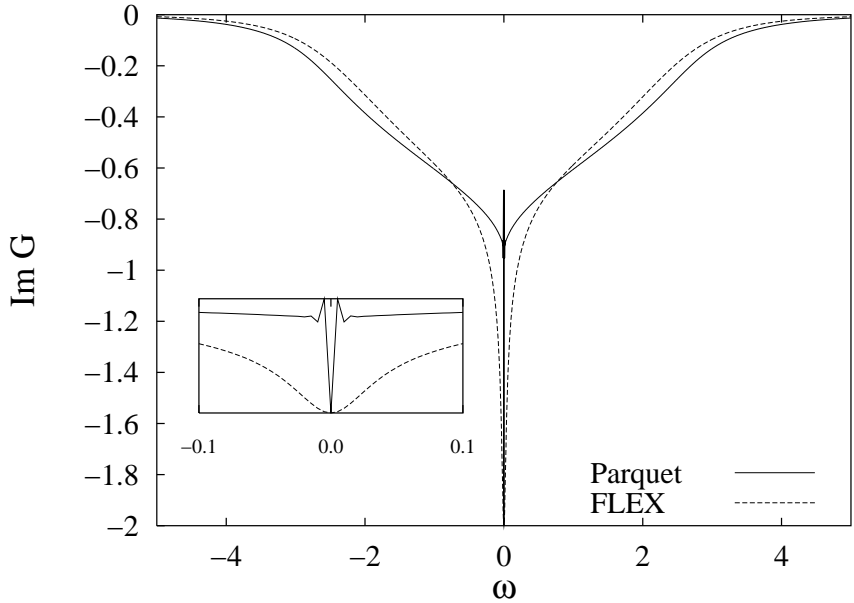,height=12cm, angle=-90}
\caption{\label{fig:FLEX-comp-ImG}Comparison of the imaginary part of the
  one-electron propagator from the FLEX and the simplified parquet
  approximations. The inset shows a tendency towards the formation of a central
  quasiparticle peak in the parquet approximation.}
\end{figure}

\begin{figure}
  \epsfig{figure=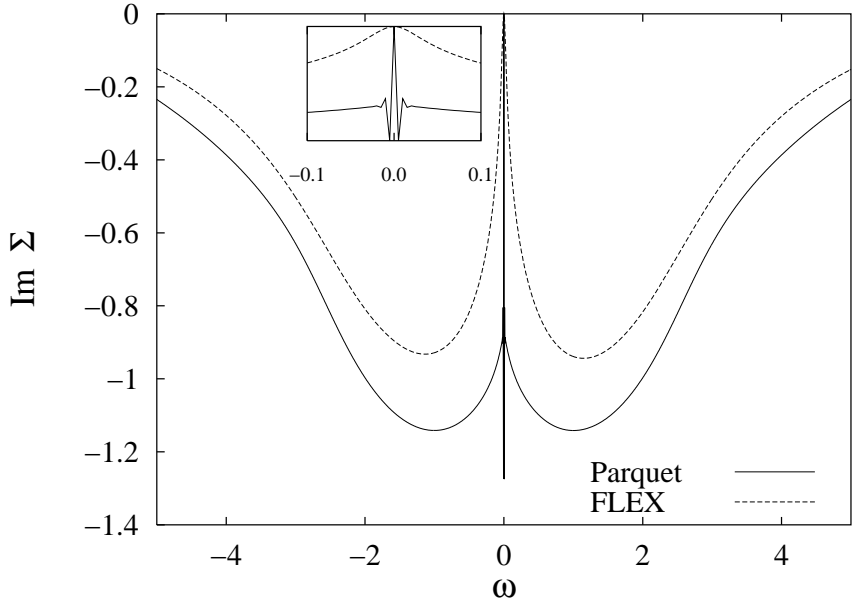,height=12cm, angle=-90}
\caption{\label{fig:FLEX-comp-ImS}Comparison of the imaginary part of the
  self-energy with the layout as in Fig.~\ref{fig:FLEX-comp-ImG}.}
\end{figure}

Although we were unable to iterate the solutions for interactions
approaching the metal-insulator transition, we nevertheless can demonstrate
a qualitative difference in the behavior of the FLEX and the parquet
approximations. Apart from the fact that the FLEX approximation becomes
unstable, the main qualitative difference is evident for small frequencies.
Fig.~\ref{fig:FLEX-comp2} shows the difference in the frequency behavior of
the real part of the vertex function $\Gamma$ at $U=2w$. A tendency to
forming of a narrow quasiparticle peak around the Fermi energy is evident
in the parquet approximation. The parquet approximation is much closer to
the instability point $\Gamma(0)=-1$ than the FLEX solution. However, due
to the self-consistence in the parquet approach, the solution remains
stable.
Fig.~\ref{fig:FLEX-comp-ImG} and Fig.~\ref{fig:FLEX-comp-ImS} show the same
for the imaginary part of the one-electron propagator and the self-energy.
A tendency to the expected insulating solution and splitting of the
quasiparticle central band at low frequencies can again be clearly
recognized. However, the low-energy dynamics in the parquet approximation
is different from the unstable IPT solution \cite{Georges96}.  The central
quasi-particle peak is much more narrow, however much less separated from
the excited states. No pronounced quasigap is yet present.  The high
frequency behavior of the density of states should be dominated by
satellite peaks.  Emerging of shoulders for the expected lower and upper
Hubbard bands can be seen in the parquet solution (but not in FLEX).  Also
the satellite bands (the high-energy region) are less pronounced than in
the IPT and numerical renormalization group \cite{Bulla98} or Monte Carlo
simulations \cite{Jarrell92,Georges96}. Hence the separation of low- and
high-energy scales seems to go much slower in the diagrammatic approaches
than in numerical simulations. It can be explained by the fact that the
spin-symmetric atomic limit with a discrete energy spectrum is not easily
reproducible in the parquet construction with a continuous spectrum of
energies. It is the half-bandwidth
$w$ that fixes the energy scale and alike the Bethe-ansatz solution it
cannot simply be limited to zero. No tendency to a divergence in the
self-energy is observed, as seen in the numerical renormalization group
\cite{Bulla98}. It is impossible to produce a singular self-energy in the
metallic phase in the Schwinger-Dyson equation with integrable two-particle
functions.

\begin{figure}
  \epsfig{figure=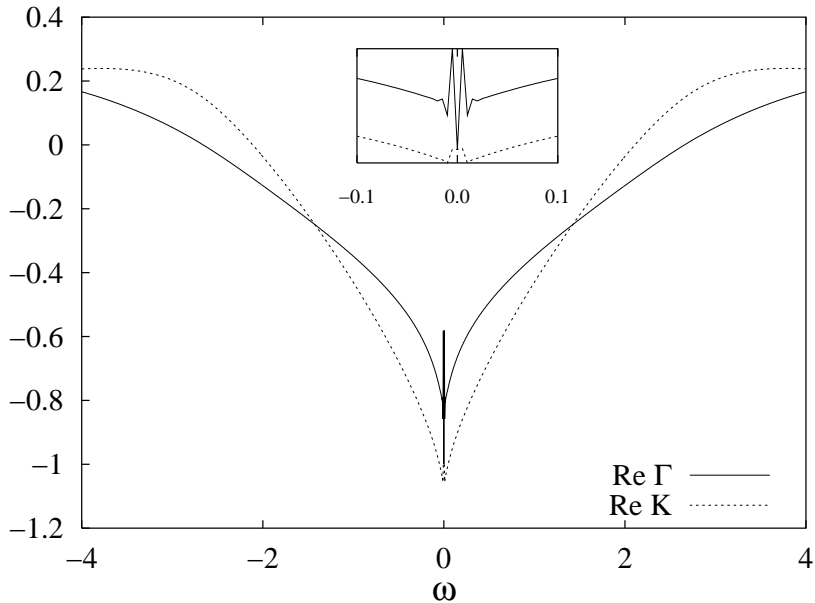,height=12cm, angle=-90}
\caption{\label{fig:parq-2P}Real parts of the two vertex functions $\Gamma$
  and ${\cal K}$ calculated with the simplified parquet approximation at
  $U=2w$.  Although these functions coincide at the Fermi energy their
   dynamical behavior is different.}
\end{figure}

The parquet approximation has two vertex functions that are to show the
same divergence. However, their dynamical behavior beyond the Fermi energy
is different as shown in Fig.~\ref{fig:parq-2P}. It is the irreducible
vertex function from the interaction channel ${\cal K}$ that has a stronger
tendency towards instability. The values of the two vertex functions at the
Fermi level coincide within the numerical resolution and lie very close to
$-1$, within the precision of $10^{-9}$.

The critical point of the metal-insulator transition is not directly
accessible with the present numerical solution of the simplified parquet
approximation. However, the low frequency behavior of the divergent
functions can be assessed in leading order analytically. At the critical
point of the metal-insulator transition both vertex functions,
$U/(1+\Gamma(\omega))$ and $U/(1+{\cal K}(\omega))$, diverge at the Fermi
energy $\omega=0$. Consequently, the derivatives of the real part of the
self-energy (effective mass) and the derivatives of the imaginary parts of
the vertex functions diverge. It is straightforward to obtain in leading
order
\begin{mathletters}\label{eq:sing}
\begin{eqnarray}
  \label{eq:K-Gam-sing}
  \mbox{Im}\ {\cal K}'_\sigma(i0^+)&\doteq& -\frac U\pi\ \frac{\left
      [ \mbox{Im}\ G_\sigma(i0^+)\right]^2}{1 +\Gamma_{-\sigma}(0)}\ ,
  \hspace{10pt}\mbox{Im}\ \Gamma'_\sigma(i0^+)\doteq -\frac U\pi\
      \frac{\mbox{Im}\ G_{\sigma}(i0^+)\mbox{Im}\ G_{-\sigma}(i0^+)} {1-{\cal
      K}_{-\sigma}(0){\cal K}_{\sigma}(0)}\ ,\\[4pt] 
  \label{eq:Sigma-sing} 
  \mbox{Re}\ \Sigma'_\sigma(0)&\doteq&-\frac U\pi\ \left[ \mbox{Im}\
  G_{-\sigma}(i0^+)\frac{X_{\sigma-\sigma}(0)}{1+\Gamma_\sigma(0)}
  +\mbox{Im}\ G_{\sigma}(i0^+)\frac{X_{-\sigma-\sigma}(0)}{1-{\cal
  K}_{-\sigma}(0){\cal K}_{\sigma}(0)} \right]  
\end{eqnarray}\end{mathletters}
where the small (Kondo) energy scale is $\Delta=w(1+\Gamma(0))$ or
equivalently $\Delta=w(1+{\cal K}(0))$. We see from (\ref{eq:sing}) that
the divergence of the effective mass is bound to a divergence in the
electron-hole dominated irreducible vertex functions.  It is important to
stress that a singularity in one channel causes a nonanalyticity in the
other channel. Hence suppressing any of the channels with electron-hole
loops in the effective impurity problems means inappropriate treatment of
the metal-insulator (Kondo) critical region.

\begin{figure}
  \epsfig{figure=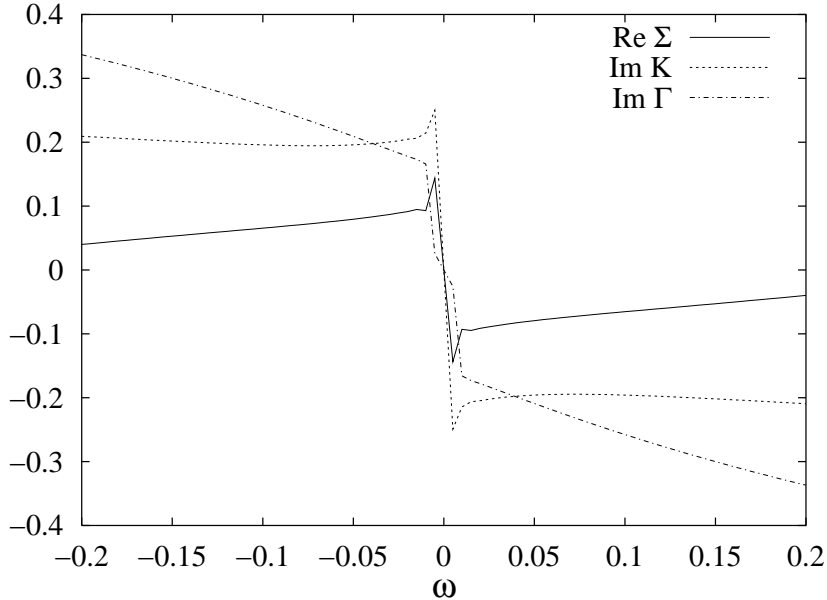,height=12cm, angle=-90}
\caption{\label{fig:parq-div} Three functions with  divergent slopes at the 
  metal-insulator transition from (\ref{eq:sing}) at $U=2w$.}
\end{figure}

If the Schwinger-Dyson equation of motion is fulfilled the effective mass
cannot diverge, or a sharp metal-insulator transition cannot appear without
a singularity in the electron-hole (local) vertex functions.
Fig.~\ref{fig:parq-div} shows for $U=2w$ the three functions the slopes of
which should be infinite at the transition point. The least sharp slope
displays the vertex function $\Gamma$ as can be expected from the more
enhanced tendency to instability in ${\cal K}$ than in $\Gamma$.

The spin-symmetric situation is the most difficult to analyze because of
the degeneracy in the spin space (singlet and triplet scatterings are
indistinguishable). If we switch on an external magnetic field we remove
the spin symmetry and the density of states will no longer be pinned at the
Fermi level.  The value of ${\cal K}_{\sigma}(0){\cal K}_{-\sigma}(0)$
decreases and we push the solution away from the criticality of the
metal-insulator transition. Going from weak to strong coupling at a fixed
field, we gradually increase the absolute value of $\Gamma_\sigma(0)$ as
long as we reach a critical point (for sufficiently small fields)
$\Gamma_\sigma(0)=-1$ at the boundary to the fully spin-polarized (Hartree)
state $U_{s}$. The value of ${\cal K}_{\sigma}(0){\cal K}_{-\sigma}(0)$
first increases with interaction, reaches a maximum at intermediate
coupling and then, due to the Hartree band-splitting terms in the
self-energy, decreases to zero at the boundary to the fully spin-polarized
state. There is no indication for a Mott-Hubbard metal-insulator transition
in an external magnetic field. An extrapolation from finite fields to $B=0$
suggests a hypothesis that the metal-insulator critical point is shifted to
infinity as in the single impurity case. However, neither finite fields do
allow for a stable numerical solution with linearly distributed mesh points
at strong coupling to support the hypothesis quantitatively.  A detailed
quantitative analysis of the Hubbard model in an external magnetic field
will be presented elsewhere.

\section{Discussion and conclusions}
\label{sec:conclusions}

We demonstrated in this paper the importance of a proper formulation of
approximate theories in order to determine the range of their
applicability.  We showed that a careful stability analysis is needed in
intermediate or strong-coupling regimes due to instabilities, poles that
may arise in two-particle functions.  We must have a reliable description
of two-particle functions to be able to decide about stability of
solutions. We suggested that the parquet approach formulated here as an
extension of the Baym-Kadanoff construction with properly chosen external
disturbances offers the desirable framework for a consistence and stability
analysis of approximate solutions. Under a physically plausible argument
that singularities in two-particle functions are  due to poles in the
Bethe-Salpeter equations, the parquet approach leads to a qualitatively
correct asymptotic critical behavior at transition points. It introduces a
new two-particle self-consistence causing that two-particle functions can
contain only integrable divergences as in the exact theory.

We found a complete set of criteria for local instabilities from the
poles in the Bethe-Salpeter equations. We attached to each singularity in
two-particle functions an order parameter breaking a symmetry of the
underlying Hamiltonian. The parameters are new self-energies and are either
of density type, such as magnetization or have anomalous character and do
not conserve either spin (transverse magnetic order) or charge
(superconductivity). Since any approximate self-consistent theory breaks
some relations between various Green functions, we always have two sets of
two-particle functions and stability criteria. The two-particle functions
determined from the Schwinger-Dyson equation of motion are used for
internal stability and consistency of the theory and determine whether
spectral functions have the desired analytical properties. The two-particle
functions defined thermodynamically from the Luttinger-Ward functional via
functional derivatives are used to determine external stability w.r.t.
external disturbances or dynamical fluctuations. The two different
two-particle functions must produce qualitatively the same phase diagram
and spectral properties in a trustworthy theory.

\begin{figure}
  \epsfig{figure=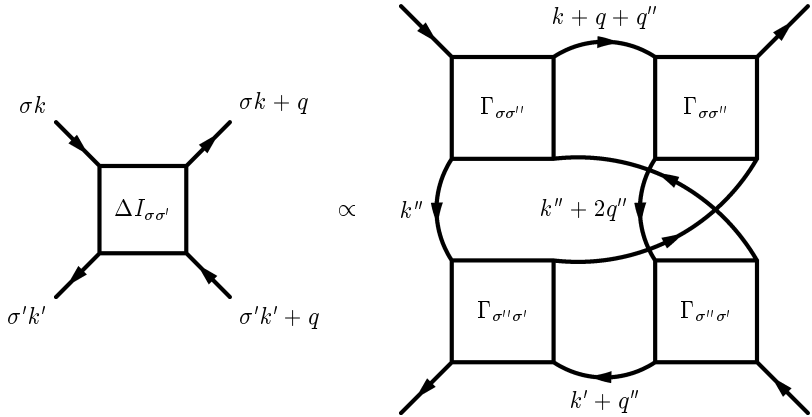,height=6cm}
\caption{\label{fig:beyond-parquet} A lowest-order contribution to the
  completely two-particle irreducible vertex beyond the parquet
  approximation. The bare interaction is replaced by the full two-particle
  vertex from the parquet solution. Double primed variables are summation
  indices.}
\end{figure}

Approximations in the Baym scheme are introduced for the full two-particle
vertex function in the Schwinger-Dyson equation. Since the vertex function
may get singular,  approximations to the Schwinger-Dyson
equation can make the resulting theory unreliable at two-particle
criticality. The presented parquet approach does not touch the
Schwinger-Dyson and Bethe-Salpeter equations of motion and introduces
approximations only in the input to these equations, namely in the
completely irreducible two-particle vertex. This is a new qualitative
change in describing critical behavior, since unlike the full vertex
function, the completely irreducible vertex is not expected to cause new
singularities to appear. Diagrammatically, it can only be broken into two
separate parts by cutting at least three one-particle propagators.  It
reduces in the parquet approximation to the bare interaction. A
lowest-order correction to it is plotted in Fig.~\ref{fig:beyond-parquet}.

Corrections to the completely irreducible vertex beyond the parquet
approximation consist of integrals over potentially singular functions and
do not generate new divergences. These corrections may, however, change the
critical exponents \cite{Bickers92}.  However, an expansion beyond the
parquet approximation must not be interchanged with a weak-coupling loop
expansion.  The parquet approximation can be viewed upon as a sophisticated
``mean-field'' theory with dynamical vertex corrections for strongly
correlated systems. Strongly correlated electrons do not allow for simpler
static mean-field theories, since two-particle criticality may
realistically be described only if we do not loose integrability of
divergences and stability in all two-particle channels. All simpler
approaches loose this essential property of the exact theory.

The most difficult problem with the parquet approach is that even the
simplest approximation does not allow for explicit nonperturbative
solutions in closed form. Hence we need further simplifications to obtain a
workable approximate scheme. We proposed such a scheme and tested it on
effective impurity models with frequencies as dynamical variables. The main
idea of our simplification is to abandon all irrelevant variables that do
not influence the existence and quality of singularities in two-particle
functions. We reduced the number of relevant two-particle channels at half
filling to two and kept only the variables at which the divergence in the
appropriate channel arises.  The resulting simplified theory is comparable
in complexity with the single-channel (FLEX) approximations, but behaves in
a qualitatively different manner at intermediate and strong coupling.  We
showed explicitly on the Hubbard and single-impurity Anderson models in an
external magnetic field that all simpler approximations get unstable at
intermediate coupling and fail to describe the transition from weak to
strong coupling. Only the parquet approximation remained stable on the
metallic side of the Mott-Hubbard metal-insulator transition.  Although
analytically stable, the simplified parquet approximation with linearly
distributed integration mesh points became numerically unstable when used
in the critical region of the Mott-Hubbard metal-insulator transition.
This instability is of purely numerical character and does not follow from
the quality of the parquet approximation itself.

A picture from the parquet theory for the transition from weak to strong
coupling in the mean-field solution of the Hubbard model has not yet been
completed, but the presented analysis strongly suggests in agreement with
\cite{Kehrein98}, that there cannot be a sharp, second-order Mott-Hubbard
transition for finite coupling at zero temperature. First, we have not
found a sharp transition when an external magnetic field is applied and the
only divergence in the two-particle spin-flip function was found at the
boundary to the fully spin-polarized solution. Second, if there were a
sharp transition in the paramagnetic, spin-symmetric phase, then inevitably
two local two-particle functions from the electron-hole and the interaction
channels had to diverge at the transition point. It follows from our
stability analysis that then a local order should set in. However, local
two-particle functions are sums of momentum-dependent two-particle
functions over the Brillouin zone, even in infinite dimensions. If such a
sum diverges, it must be at least one momentum ${\bf q}_0$ at which the
paramagnetic solution gets unstable before the Mott-Hubbard transition is
reached.  Hence, a long-range order will conceal the Mott-Hubbard
metal-insulator critical point. We conclude that Fermi liquid can break
down only if a long-range order sets in.

Based on the experience and understanding of the origin of singularities at
intermediate and strong coupling, we can conclude that only theories with
dynamical vertex renormalizations are capable to describe the transition
from weak to strong coupling qualitatively correctly.  The parquet
approximation with an adequate numerical technique to iterate or
diagonalize the resulting nonlinear equations offers a suitable
analytic-numerical tool for this purpose. Especially quantum criticality is
the situation where only the parquet approach is able to give a complete
picture of the critical behavior and symmetry-breaking order parameters.
What is to be done in particular cases is to find appropriate
simplifications making the parquet theory viable but retaining the
essential physics.

To summarize, we have presented a general parquet construction treating
one- and two-particle Green functions on the same footing within a single
self-consistent renormalization scheme. We showed that instabilities are
induced by divergences in two-particle functions to which symmetry-breaking
order parameters are always attributed, either normal of density type or
anomalous.  Further on, an instability in an approximate theory need not
definitely indicate a transition to a long-range order. A singularity in
one two-particle channel induces singular vertex corrections
in the other channels that tend to suppress the long-range order. Even if a
divergence appears only in one channel, as in an external magnetic field, a
singularity in the other channels is generated and at least one another
channel is needed to introduce a two-particle self-consistence and to keep
the solution stable. The parquet approach is always to be used where
quantum phase transitions are present and stability criteria in at least
two channels may be broken.  It is the low-temperature limit in low
spatial dimensions ($d=0,1,2$) and in intermediate- and strong-coupling
regimes in three and higher dimensions. There, the parquet approximation is
the simplest consistent and qualitatively correct description.

The work was supported by grant No. 202/98/1290 of the Grant Agency of the
Czech Republic.

\begin{appendix}
  \setcounter{equation}{0} \renewcommand{\thesection}{}
  \renewcommand{\theequation}{\Alph{section}.\arabic{equation}}
\section{}
To derive the generating thermodynamic functional for the parquet
approximation we proceed as generally outlined in \cite{Dominicis62} and
explicitly evaluated for the parquet approximation in \cite{Janis98b}.
I.~e. we introduce pairwise Legendre conjugate variables for the
irreducible and reducible two-particle vertex functions. We then integrate
the Bethe-Salpeter equations (\ref{eq:Bethe-Salpeter}) so that
\begin{eqnarray}
  \label{eq:Phi-dif}
  {\cal K}^\alpha=\frac{\delta\Phi_\Lambda}{\delta\Lambda^{\alpha}} \ ,
  &\hspace{1cm}& \Lambda^\alpha= \frac{\delta\Phi_K}{\delta{\cal K}^
    {\alpha}}\ .
\end{eqnarray}
We use the notation introduced in Sec.~\ref{sec:stability} and particularly
(\ref{eq:channels}) and write the functionals as traces over the active
variables in each channel
\begin{eqnarray}
  \label{eq:Phi-U}
  \Phi^{U}_\Lambda&=&\frac 1{2\beta{\cal N}}\sum_{q}\mbox{Tr}_{\sigma k}
   \left\{\left[\Lambda^{U}G^{(2)}\right]^U[q]-\frac 12\left[\Lambda^{U}
   G^{(2)}\right]^U[q]\star\left[\Lambda^{U}G^{(2)}\right]^U[q]\right.
   \nonumber \\ &&\hspace*{60pt}
  \left. -\ln_\star\left(1+\left[\Lambda^{U}G^{(2)}\right][q]\right)
   \right\} \ ,  
\end{eqnarray}
\begin{eqnarray}
  \label{eq:Phi-eh}
  \Phi^{eh}_\Lambda&=&\frac 1{2\beta{\cal N}}\sum_{\sigma\sigma' q}\mbox{Tr}_{k}
    \left\{\left[\Lambda^{eh}G^{(2)}\right]^{eh}_{\sigma\sigma'}[q] -\frac
    12\left[\Lambda^{eh} G^{(2)}\right]^{eh}_{\sigma\sigma'}[q]\bullet
    \left[\Lambda^{eh}G^{(2)}\right]^{eh}_{\sigma\sigma'}[q]\right.
    \nonumber \\ &&\hspace*{60pt}
   \left. -\ln_\bullet\left(1+\left[\Lambda^{eh}G^{(2)}\right]^{eh}
    _{\sigma\sigma'}[q]\right)\right\} \ , 
\end{eqnarray}
\begin{eqnarray}
  \label{eq:Phi-ee}
  \Phi^{ee}_\Lambda&=&\frac 1{2\beta{\cal N}}\sum_{\sigma\sigma' q}\mbox{Tr}_{k}
    \left\{\left[\Lambda^{ee}G^{(2)}\right]^{ee}_{\sigma\sigma'}[q] -\frac
    12\left[\Lambda^{ee} G^{(2)}\right]^{ee}_{\sigma\sigma'}[q]\circ
    \left[\Lambda^{ee}G^{(2)}\right]^{ee}_{\sigma\sigma'}[q]\right.
    \nonumber \\  &&\hspace*{60pt}
   \left. -\ln_\circ\left(1+\left[\Lambda^{ee}G^{(2)}\right]^{ee}_{\sigma
    \sigma'}[q]\right)\right\} \ 
\end{eqnarray}
where $G^{(2)}$ is the free two-particle propagator defined in
(\ref{eq:parquet-G2}). Note that the trace in the interaction channel
includes spin as an active variable. The logarithms with subscript denote
the appropriate matrix multiplication used in their power series
expansions.

Integration of the other set of two-particle equation of the parquet
approximation leads to a functional
\begin{eqnarray}
  \label{eq:Phi-K}
  \Phi_K&=&\frac 1{2\beta{\cal N}}\sum_{\sigma\sigma' q}\mbox{Tr}_{k}\left\{ 
   \left[\left(U+{\cal K}^{ee}\right)G^{(2)}\right]^{eh}_{\sigma\sigma'}[q]
   \bullet\left[{\cal K}^{eh}G^{(2)}\right]^{eh}_{\sigma\sigma'}[q]
   +\left[UG^{(2)}\right]^{ee}_{\sigma\sigma'}[q]\circ\left[{\cal K}^{ee}
   G^{(2)}\right]^{ee}_{\sigma\sigma'}[q]\right\}
 \nonumber\\ &&
   + \frac 1{2\beta{\cal N}}\sum_{q}\mbox{Tr}_{\sigma k}\left\{\left[
   \left(U+{\cal K}^{eh}+{\cal K}^{ee}\right)G^{(2)}\right]^U[q]\star
   \left[{\cal K}^U G^{(2)}\right]^U[q]\right\} \ .
\end{eqnarray}
Be aware that the operator of the bare interaction acts only between
different spins, i.~e. contains $\delta_{\sigma,-\sigma}$.  The functionals
$\Phi^\alpha_\Lambda$ are exact and only the functional $\Phi_K$ is
approximate, since only it depends on the completely irreducible vertex
being replaced here by the bare interaction.

The Luttinger-Ward generating functional is a sum of $\Phi^\alpha_\Lambda$
and $\Phi_K$ completed with second-order term and products of the Legendre
conjugate two-particle functions $\Lambda$ and ${\cal K}$ to keep the
variation of the Luttinger-Ward functional w.r.t. two-particle functions
zero.  We obtain
\begin{eqnarray}
  \label{eq:Phi-full}
  \Phi[G;\Lambda,{\cal K};U]&=&\frac 1{2\beta{\cal N}}\sum_{q}\mbox{Tr}
      _{\sigma k}\left\{\left[\Lambda^U{G}^{(2)}\right]^U[q]\star\left[
      {\cal K}^U G^{(2)}\right]^U[q]-\frac 12\left[UG^{(2)}\right]^U[q]
      \star\left[UG^{(2)}\right]^U[q]\right\}\nonumber\\    
  &&\hspace*{-50pt}+\frac 1{2\beta{\cal N}}\sum_{\sigma\sigma' q}\mbox{Tr}_{k}
      \left\{\left[\Lambda^{eh}G^{(2)}\right]^{eh}_{\sigma\sigma'}\bullet
      \left[{\cal K}^{eh}G^{(2)}\right]^{eh}_{\sigma\sigma'}[q]+\left[
      \Lambda^{ee}G^{(2)}\right]^{ee}_{\sigma\sigma'}[q]\circ\left[
      \Lambda^{ee}G^{(2)}\right]^{ee}_{\sigma\sigma'}[q]\right\}
      \nonumber\\[4pt] 
   &&-\Phi_K\left[G;{\cal K};U\right]-\Phi^{eh}_\Lambda\left[G;\Lambda^{eh} 
  \right]-\Phi^{ee}_\Lambda\left[G;\Lambda^{ee}\right]-\Phi^{U}_\Lambda
      \left[G;\Lambda^{U}\right]
\end{eqnarray}
where we introduced in each functional its explicit dependence on the
variational one- and two-particle functions and the bare interaction $U$.

The grand potential generating the complete thermodynamics of the parquet
approximation is constructed from the Luttinger-Ward functional and a
free-electron term with the Hartree contribution \cite{Baym62}. We hence
add also the particle densities as additional variational parameters and
obtain for a fixed $\mu_\sigma=\mu+\sigma B$
\begin{eqnarray}
\label{eq:Omega}
 \frac 1{{\cal N}}\Omega [n_{\uparrow},n_{\downarrow};\Sigma ,G;\Lambda,
 {\cal K}]&=&-\frac 1{\beta 
   {\cal N}}\sum_{\sigma n,{\bf k}}e^{i\omega_n0^{+}}\Big\{ \ln \left[
     i\omega _n+\mu _\sigma -\epsilon({\bf k}) -Un_{-\sigma }-\Sigma_\sigma
     ({\bf k},i\omega_n)\right] \nonumber\\[4pt]         
& & \hspace*{10pt} +G_\sigma ({\bf k},i\omega_n)\Sigma _\sigma ({\bf
    k},i\omega_n)\Big\} -Un_{\uparrow}n_{\downarrow} +\Phi [G;\Lambda,
{\cal K};U] \ .
\end{eqnarray}
Here $n_{\uparrow},n_{\downarrow};\Sigma ,G;\Lambda,{\cal K}$ are
independent variables and complex functions. Their physical values are
chosen from stationarity of the grand potential (\ref{eq:Omega}) with
respect to variations of its independent variables/functions.  There are
three pairs of Legendre conjugate variational variables in
(\ref{eq:Omega}). The Hartree parameters $n_\uparrow$ and $n_\downarrow$,
the one-electron functions $\Sigma_\sigma(k)$ and $G_\sigma(k)$, and
finally the two-particle irreducible and reducible functions
$\Lambda^\alpha_{\sigma \sigma'}(k,k';q)$ and ${\cal
  K}^\alpha_{\sigma\sigma'}(k,k';q)$, respectively.

\end{appendix}



\begin{references}
\bibitem{note0} The interaction channel is alternatively called crossed
  electron-hole channel. See Refs.~\cite{Weiner70,Bickers91,Janis98a} for
  the nomenclature concerning the two-particle irreducibility in the
  Hubbard model.
\bibitem{note1} The channel dependence in the vertex function $\Gamma$
  reduces to a mere rotation or interchange of incoming and outgoing lines.
  We can drop the superscript $\alpha$ when fixing the incoming and
  outgoing particles as in Fig.~\ref{fig:2P-generic}.
\bibitem{note2} Our convergence criterion was that integrals over the
  square of the differences between subsequent iterations fall below
  $10^{-3}$.
\end{references}
\end{document}